**Role of Jet Spacing and Strut Geometry on the formation of Large Scale Structures & Mixing Characteristics**

Rahul Kumar Soni, Ashoke De [a)]

*Department of Aerospace Engineering, Indian Institute of Technology Kanpur, Kanpur, 208016, India*

The present study primarily focuses on the effect of jet spacing and strut geometry on the evolution and structure of the large-scale vortices which play a key role in mixing characteristics in turbulent supersonic flows. Numerically simulated results corresponding to varying parameters such as strut geometry and jet spacing ($X_n=nD_j$ such that $n=2, 3$ & $5$) for square jet of height $D_j=0.6$ mm is presented in the current study, while the work also investigates the presence of the local quasi-two-dimensionality for the $X_2(2D_j)$ jet spacing; however, the same is not true for higher jet spacing. Further, the tapered strut (*TS*) section is modified to the straight strut (*SS*) for investigation, where the remarkable difference in flow physics is unfolded between the two configurations for similar jet spacing ($X_2: 2D_j$). The instantaneous density and vorticity contours reveal the structures of varying scales undergoing different evolution for the different configurations. The effect of local spanwise rollers is clearly manifested in the mixing efficiency and the jet spreading rate. *SS* configuration exhibits excellent near field mixing behavior amongst all the arrangements. However, in case of *TS* cases, only $X_2(2D_j)$ configuration performs better due to the presence of local spanwise rollers. The qualitative and quantitative analysis reveals that near-field mixing is strongly affected by the two-dimensional rollers, while the early onset of wake mode is another crucial parameter to have improved mixing. Modal decomposition performed for *SS* arrangement sheds light into the spatial and temporal coherence of the structures, where the most dominant structures are found to be the *Von-Karman street* vortices in the wake region.

**I. INTRODUCTION**

The successful design of Scramjet engine requires efficient combustor design among many other design challenges. An efficient combustor would mean the residence time for the fuel-air mixture is modified in such a manner that mixing time scale is reduced significantly to offer better combustion. Also mixing length scale, i.e. the near-field mixing characteristics seem to be very important. It is essential to have a better understanding regarding the mixing behavior of two high speed (supersonic) streams. Various studies[1-4] performed in the past especially with splitter plate put forward that for high-speed streams mixing are inhibited due to the presence of the three-dimensionality. Brown & Roshko[2] and Bogdanoff[4] pointed out that the role of *K-H* instability is limited in high convective Mach number and this is manifested in the overall physics. It is well known that *K-H* instability gets triggered across the shear layer due to the high-velocity gradient, thereby leading to the formation of large spanwise roller mostly quasi-two-dimensional in nature. Clemens & Mungal[5,6] carried out an experiment over the range of convective Mach numbers ranging from low incompressible values to higher supersonic values. They noticed more organized structures mostly two dimensional for low convective Mach number, while for higher supersonic values lack of coherence and presence of three-dimensionality was noted. Barre et al.[7] from their experimental work concluded that the turbulent diffusion in free shear flow depends mainly on the large-scale structures. Soni & De[8] in their recent work involving planar jet in strut configuration also concluded that formation of large-scale structures is indeed very critical for the overall mixing process. However, contradictory to popular belief they also reported that the increase in the convective Mach number leads to the increased mixing layer thickness for supersonic convective Mach number. They also pointed out that strength and unsteadiness of the recirculating region at the strut base also play a key role in the formation of large-scale vortices.

From literature, one realizes that there exist various injection techniques and one may refer to the work of Gutmark et al.[9] and Seiner et al.[10] for detailed review. They proposed that passive injection techniques that are capable of generating and introducing low frequency streamwise vortical structures are most effective at supersonic speeds. Parallel introduction of the injectant in the oncoming supersonic flow is one such example where noticeable increment in the mixing characteristics has been observed. An additional perk of parallel injection is reflected in the contribution towards the thrust and low total pressure loss compared to normal injection. The shock wave generated at the wedge/strut tip appear to be lucrative as they lead to the shear layer instability through shock/shear layer interaction (*SSLI*) but one must be careful about the shock strength. Interaction of the shock wave and the shear layer leads to the production of vorticity due to the non-alignment of pressure and density gradients which leads to enhanced strain rate and results in better mixing through the turbulent diffusion. Several researchers[11-14] also proposed that the presence of the wake flow around the base region is also responsible for the exchange of the mass and momentum across the shear layer. Experimental studies by Drummond & Givi[15] reported that the combustion and mixing efficiency was significantly reduced in the absence of the shock wave. However, the study of Huh & Driscoll[16] showed that the flame stability limit was improved when oblique shock wave of optimum strength was introduced. Budzinsky et al.[17] conducted the experiment to characterize the effect of normal shock on mixing and reported at least two-fold increment in molecular level mixing for a single interaction, while it was around three times for the double interaction case.

---

[a)] Author to whom correspondence should be addressed: ashoke@iitk.ac.in

Gerlinger and Bruggemann[11, 18], Yu et al.[19] reported in their study that the size and shape of large-scale structures are strongly dependent on the finite lip thickness at the strut base. Zhang et al.[20, 21] carried out the experimental exercise to establish the role of vortices in the mixing of high-speed flows. They utilized lobed mixer with peak and trough configuration and reported that the large-scale structures shedding from the peak and trough region play vital role in the mixing process. They also reported that interaction of streamwise structures with the *K-H* vortices leads to the formation of large spanwise structures which positively affects the momentum transfer between the two streams.

Therefore, in the light of above discussion, we performed Large Eddy Simulation (*LES*) of two strut configurations namely Straight strut (*SS*) and Tapered Strut (*TS*) with three injectors located at the base region for different injector spacing. The main focus of the present investigation is three fold i) to verify and characterize the effect of spanwise separation of injectors on the overall mixing process for *TS* configurations, ii) to study the same but for altered configuration, i.e. *SS* arrangement for $X_2(2D_j)$ spacing, and iii) finally perform the modal decomposition to separate the various modes present in such flows to characterize mixing. Essentially in the first part, we attempt to understand the manifestation of the three-dimensionality in the mixing due to the presence of the mostly streamwise vortical structures; and in the second part, we report the detailed flow physics by altering the formation of structures and study its role by comparing the $X_2$-*TS*$(2D_j)$ and $X_2$-*SS*$(2D_j)$. In this whole exercise, the convective Mach number ($M_c$) has been maintained same for all cases, i.e. all results reported herein correspond to $M_c = 1.4$. So, this leads to the curiosity whether higher convective Mach number is truly a strong parameter in the context of three-dimensionality or there are other parameters that may play a significant role in modifying the mixing characteristics. The loss of coherence in the flow field at higher convective Mach number inhibits the mixing layer growth[6]. Recently, the previous work by the authors[8] involving tapered & straight strut with planar jet injection reported that convective Mach number is not the strong parameter that affects the mixing. They concluded that high velocity gradient and turbulence amplification across the shear layer leads to the formation of the large scale structures which positively alter the turbulent diffusion and thereby the mixing. However, in the present research, the planar jet is replaced with the multi-jet (square jet) configuration where jet spacing [$X_2(2D_j)$, $X_3(3D_j)$ & $X_5(5D_j)$] is varied for *TS* arrangement and modified configuration, i.e. *SS* case is reported for $X_2(2D_j)$ jet spacing only. It is expected that the variation in the strut geometry will induce different recirculation region whose effect will be reflected in the mixing performance.

We attempt to explore the nature of flow for such configuration, as it is expected that the physics governing the mixing process will be affected by the jet spacing. Since it is established from the literature that the three-dimensionality in case of splitter plate configuration inhibits the mixing due to lack of coherence in structures, it will be interesting to subject real Scramjet injector to similar test for high convective Mach number where compressibility effects are dominating. Further, we deploy modal decomposition techniques namely Proper Orthogonal Decomposition (POD) and Dynamic Mode Decomposition (DMD) to get an insight into the spatial and temporal coherence structures dominating the flow physics.

**II. Numerical Details**

The Favre-filtered governing equations for the conservation of mass, momentum, energy, and species are solved in compressible form and are given as:

Continuity equation:
$$\frac{\partial}{\partial t}(\bar{\rho}) + \nabla \cdot \left( \bar{\rho} \tilde{\vec{V}} \right) = 0 \tag{1}$$

Momentum equation:
$$\frac{\partial}{\partial t}(\bar{\rho}\tilde{\vec{V}}) + \nabla \cdot \left( \bar{\rho} \tilde{\vec{V}} \tilde{\vec{V}} \right) = -\nabla \cdot (\tilde{p}I) + \nabla^2 \left( (\mu + \mu_t) \tilde{\vec{V}} \right) \tag{2}$$

Energy equation:
$$\frac{\partial}{\partial t}(\bar{\rho}\tilde{E}) + \nabla \cdot \left( \bar{\rho} \tilde{\vec{V}} \tilde{E} \right) = -\nabla \cdot \left( (-\tilde{p}I + \mu \nabla \tilde{\vec{V}}) \tilde{\vec{V}} \right) + \nabla^2 \left( \left( k + \frac{\mu_t C_p}{\Pr_t} \right) \tilde{T} \right) \tag{3}$$

Species Transport:
$$\frac{\partial}{\partial t}(\bar{\rho}\tilde{Y}_k) + \nabla \cdot \left( \bar{\rho} \vec{V} \tilde{Y}_k \right) = -\nabla^2 \left( \left( \bar{\rho} D_k + \frac{\bar{\rho} v_t}{Sc_t} \right) \tilde{Y}_k \right) \tag{4}$$

To perform the large eddy simulation eq. (1) – (4) is filtered by applying a low-pass filter which ensures that large eddies above certain cutoff frequencies are resolved. The filtered quantities denoted by (~) are defined as

$$\tilde{f}(x) = \int_\Omega f(x') G(x, x', \Delta) dx' \tag{5}$$

Where $\Omega$ is the domain, $G$ is the convolution kernel which is sufficiently smooth and $\Delta$ is the filter width, while ($\sim$) and (-) in above equations refer to filtered and Favre filtered quantity. Here $\rho$ is the density, $u_i$ is the velocity vector, $p$ is the pressure, $E = e + u_i^2/2$ is the total energy, where $e = h - p/\rho$ is the internal energy and $h$ is enthalpy. The fluid properties $\mu$, $D$ and $k$ are respectively the viscosity, mass diffusivity and the thermal conductivity, while $\mu_t$, $Sc_t$ and $Pr_t$ are the turbulent eddy viscosity, turbulent Schmidt number and turbulent Prandtl number respectively. The turbulent Prandtl and Schmidt numbers are defined as *0.9 & 0.5* and the dynamic viscosity is evaluated through the Sutherland's law,

$$\mu = \frac{A_s \sqrt{T}}{1 + \frac{T_s}{T}} \qquad (6)$$

Here $A_s$ is the Sutherland coefficient and $T_s$ is Sutherland temperature, the specific heat at constant pressure $c_p$ is evaluated through the following polynomial

$$c_p = R\left(\left(\left(a_4 T + a_3\right)T + a_2\right)T + a_1 T\right) + a_0\right) \qquad (7)$$

Additionally, $a_5$ & $a_6$ are the constants of integration at low and high temperature end for evaluation of enthalpy and entropy. Finally, the above sets of governing equations are closed by including the ideal gas equation.

The turbulence is modeled through the large eddy simulation approach by filtering the instantaneous governing equations. In the present study dynamic model based on two model coefficients is invoked. The filtering approach results in additional viscous term representing the sub-grid scale flow which is closed as $\tau_{ij}^{sgs} - \frac{1}{3}\delta_{ij}\tau_{kk} = -2\bar{\rho}C\bar{\Delta}^2(\tilde{S}_{ij} - \frac{1}{3}\delta_{ij}\tilde{S}_{ij})$ where $S_{ij}$ represents the strain rate. The $C = \frac{\langle L_{ij}L_{ij}\rangle}{\langle M_{kl}M_{kl}\rangle}$ & $C_I = \frac{\langle L_{kk}\rangle}{\langle \beta-\alpha\rangle}$ are the model coefficients evaluated dynamically by averaging locally further details about the model and wall modelling approach used herein can be referred to the work of Soni et al.[8, 22, 23].

### A. Proper Orthogonal Decomposition (POD) and Dynamic Mode Decomposition (DMD)

Proper orthogonal decomposition is a methodology that aids in decomposing the flow field in time and space. The modes are extracted by optimizing the mean square of the field variable in question. The POD is a very powerful technique in a sense if considering a turbulent flow field which has infinite scales and hence the infinite degree of freedom it becomes impossible to characterize each and every scale. It is well known fact that although highly chaotic and random in nature turbulent flow does exhibit certain pattern (or coherence) which can be characterized to understand the turbulent flows. Lumley[24] first introduced this technique in the study of turbulence to extract the coherent or low frequency structures. The Eigen modes are computed by collecting sufficient number of snapshots separated properly in time as temporal separation could play a vital role in the decomposition process. The decomposition is performed in such a manner that only few basis functions are able to represent the most of the energy. Basically, this allows only first few modes to represent the flow field accurately.

Although POD is a powerful tool to inspect the coherence in a flow but it lacks the temporal coherence because it is based on the second order correlation. The POD modes mostly possess multiple frequencies and being arranged on the basis of the energy content across modes, the dynamical information is lost. To remedy this, Dynamic Mode Decomposition (DMD), another class of modal decomposition techniques proposed by Schmid[25] is utilized herein. The DMD arranges the modes on the basis of pure frequencies. DMD offers certain advantages in a sense that it does not require a priori knowledge of the underlying dynamics and being temporally orthogonal pure frequencies are revealed. DMD can be thought of as enjoying both the advantages of the POD and discrete Fourier transform offering spatio-temporal coherence. More detailed information regarding POD and DMD methodology utilized herein can be found in the literature[8].

### B. Computational Details and Validation

The computational exercise is carried out in the OpenFOAM framework, a density based solver has been extensively validated by this group and published [8, 22, 26], where the existing solver is modified to include species transport equation along with the mass, momentum, and energy. The modified solver has also been extensively tested and reported by Soni & De[8]. The solver is based on the central scheme offering an alternate approach to Riemann solver with an accurate non-oscillatory solution. One may refer to the work of Greenshields et al.[27] and Kurganov & Tadmor[28] for further detail about the original density based solver. The *dynamic sgs* model invoked in the current computation is based on the two model constants which are evaluated dynamically[25]. In present simulation, second order backward Euler scheme is utilized for the time integration whereas viscid and inviscid fluxes are discretized using central difference and TVD scheme, respectively. The parallel processing is achieved through the message passing interface (MPI) technique. The statistics are collected for the simulations over ~100 non-dimensional times while maintaining CFL number below 0.5.

The computational domain as shown in Figure 1 involves a channel within which sits the strut and the coordinate system originates at the strut tip. On the backside (base region) of the strut, three injectors (colored in red) are placed such that $n=2, 3$ &5 (Figure 1) and the hydrogen is injected in the oncoming supersonic flow through these injectors. As one can see that there is a remarkable difference in the order of channel length and jet height ($D_j=0.6$ mm), multi-block meshing philosophy is adapted to maintain optimum grid size while maintaining aspect ratio. It is assured that the region prone to high gradients are accurately resolved; also care has been taken to resolve the jet region. The grid generated herein is the result of the grid independence study performed in the similar configuration with the planar jet. Since the configuration investigated herein only inherits the channel and injector dimension but the injection strategy is varied from the previous study[8]. In the present investigation square injector instead of the planar injector is employed; hence to gain confidence on the grid, some results are reported for the planar jet case only. In Figure 2, Reynolds stress and mean streamwise velocity and mean hydrogen mole fraction is presented to demonstrate the grid independence for the planar jet case in *TS* configuration, the three grids are related as *G1 (coarse) < G2 (medium) < G3 (fine)*. Overall the node distribution for *G1, G2* and *G3* case is $730 \times 142 \times 18$, $1105 \times 210 \times 25$ & $1326 \times 250 \times 30$, respectively[8]. The concerned region, i.e. jet region including the strut boundary, is resolved sufficiently to capture the shear layer and recirculation region effectively. Since the *G2 & G3* do not show any noticeable variation, in the present study the entire grids are generated similar to the *G2* as tabulated in Table I.

Table I. Grid details for the entire configuration. $y_w$ and $\Delta y_{jet}$ denote the grid spacing along near wall and jet region. $N_x$, $N_y$, and $N_z$ denote the grid point distribution along the streamwise, transverse and spanwise direction respectively, $D_j$ (=0.6 mm) signifies jet height and $N$ the total number of finite volume cells

|  | $X_2$-TS ($2D_j$) | $X_3$-TS ($3D_j$) | $X_5$-TS ($5D_j$) | $X_2$-SS ($2D_j$) |
|---|---|---|---|---|
| $\Delta y_w$ | $0.2\ D_j$ | $0.2\ D_j$ | $0.2\ D_j$ | $0.2\ D_j$ |
| $\Delta y_{jet}$ | $0.0125\ D_j$ | $0.0125\ D_j$ | $0.0125\ D_j$ | $0.0125\ D_j$ |
| $N_x$ | 1180 | 1000 | 1000 | 1180 |
| $N_y$ | 232 | 190 | 188 | 230 |
| $N_z$ | 25 | 34 | 38 | 25 |
| $N$ | $6.844\times10^6$ | $6.46\times10^6$ | $7.144\times10^6$ | $6.785\times10^6$ |

TABLE II. Inlet boundary condition for all the cases investigated and reported herein. (*M, P, U,* and *T* represent the Mach number, pressure, streamwise velocity and temperature respectively)

| Parameter | Air(Co-flow) | Hydrogen (Jet) |
|---|---|---|
| $P(kPa)$ | 49.5 | 29.5 |
| $T(K)$ | 159 | 151 |
| $M$ | 2 | 2.3 |
| $U\ (m/s)$ | 505 | 2203 |
| $Y_{N_2}$ | 0.76699 | 0 |
| $Y_{O_2}$ | 0.23301 | 0 |
| $Y_{H_2}$ | 0 | 1 |

The details related to the grid spacing and node number distribution in the different region are collated in Table I and in Table II inlet boundary condition used in the present work is reported. The jet region including the near field region, enough resolution along the streamwise and spanwise directions is ensured to properly capture the evolution of shear layer and recirculation region effectively. In the near field region, as the top and bottom walls may have very little or no influence on the mixing characteristics, these are modeled ($y+ \approx 30$) while the adopted approach has already been validated and reported in Ref.[22]. For both air and hydrogen inlet, dirichlet boundary condition is used; while the non-reflecting boundary condition is imposed at the outlet to avoid wave reflection within the domain[8,22,26]. All the wall including the top, bottom and strut region adiabatic no-slip boundary condition is prescribed and care has been taken to resolve the boundary layer. Periodicity is imposed along the spanwise boundaries, while a turbulent inlet condition is imposed at the jet. In the present case, uniform velocity is prescribed at the Air inlet because the flow upon interacting with the strut base will lead to the production of turbulence and the presence of baroclinic torque will assist in amplifying the same. It is also known that turbulence will be generated in the free shear region, especially in the present case where the velocity gradient is very high. Therefore, the downstream of the strut injector, sufficient turbulence will be generated due to above-mentioned interaction. Further, one can notice that the mixing takes place some distance from the injector, i.e. in the wake region, the flow filed is found to be sufficiently turbulent. To further assure this, the turbulent inlet is prescribed at the jet exit. Genin & Menon[29] in their study pertaining to the DLR combustor similarly imposed uniform inlet. In the previous work[8], the authors too imposed uniform condition at the inlet in the study pertaining to supersonic planar jet. Since the primary aim of this study is to look at the mixing characteristics at the downstream of the strut injector, the turbulent at the air inlet does not seem to have a significant impact on the mixing characteristics as sufficient turbulent is generated upon interaction of air stream with the strut base.

In order to check the resolution of the newly generated grids using multiple injectors, the index of quality for large eddy simulation ($LES_{IQ}$) proposed by Celik et al.[30] is presented in Figure 3; where they suggested that index of quality greater than 80% indicates sufficient grid resolution. In the present calculation, it is observed that $LES_{IQ}$ is around 95% and above and hence it can be assumed that the grid is sufficiently resolved in the entire region. Worth mentioning is the resolution of the region prone to high gradients and jet region where grid seems to perform brilliantly. Hence the grid spacing and node number distribution in the different region, as depicted in Table I, are found to be satisfactory for detailed studies.

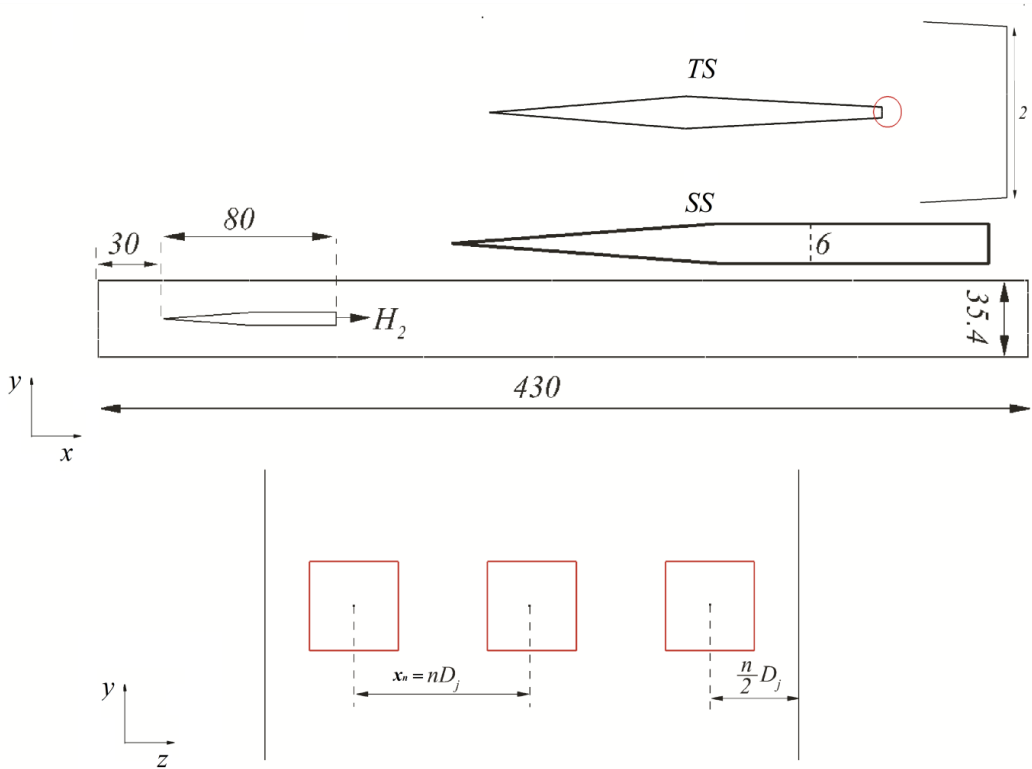

Figure 1: Details of the computational domain; all the dimensions are in mm (injectors are coloured in red)

**II. Results & Discussion**

**A. Tapered Strut (*TS*): Effect of jet spacing**

In the study pertaining to mixing characterization, it becomes imperative to get insight into the near field behavior. It is with this objective spatial variation of normalized mixing layer thickness (half thickness precisely) and mixing efficiency is presented in Figures 4 (a) & (b), respectively. In the present study, mixing layer thickness is defined as the transverse distance from the jet centerline where the hydrogen mole fraction decays to *1%* of the injected value. The mixing layer thickness is presented to appreciate the diffusion of species in the other directions and hence there is no fixed rule on the choice of the percentage value of injected species. Gerlinger & Bruggemann[11] and Ben-Yakar et al.[31] utilized 5% of injected values as the criteria to define the mixing layer thickness in their study, whereas Fuller et al.[32] and Kim et al.[33] in their study utilized 0.5% of the injected value to demonstrate the mixing layer thickness. The mixing layer thickness trend presented in Figure 4(a) points toward the similarity in profile for all configurations. However, for all the configurations studied herein, a significant difference is witnessed in the spatial growth. The various peaks present along the line plot represent the intersection of shock wave whose strength diminishes further downstream due to multiple reflections from the wall. The region close to the injector where the sudden rise is noticed represents the edge of the region where turbulent diffusion dominates the mixing of the two layers. At around $x/D_j = 334$ the profile appears to be lesser steep and grows gradually beyond this point; this signifies the edge of the high diffusive region. This suggests that in the near region especially around $x/D_j = 134$ -$266$ lot of physics is involved probably the presence of various coherent structures and difference in the shear strength ($R = u_2/u_1$, subscript 1 & 2 refer to outer flow and jet respectively) is responsible which will be explored in the subsequent section. The most striking difference that is revealed from this plot lies in the variation in values amongst different configurations, especially for the $X_2(2D_j)$ case. This clearly suggests that the spanwise separation between the two jets is reflected remarkably on the overall mixing characteristics.

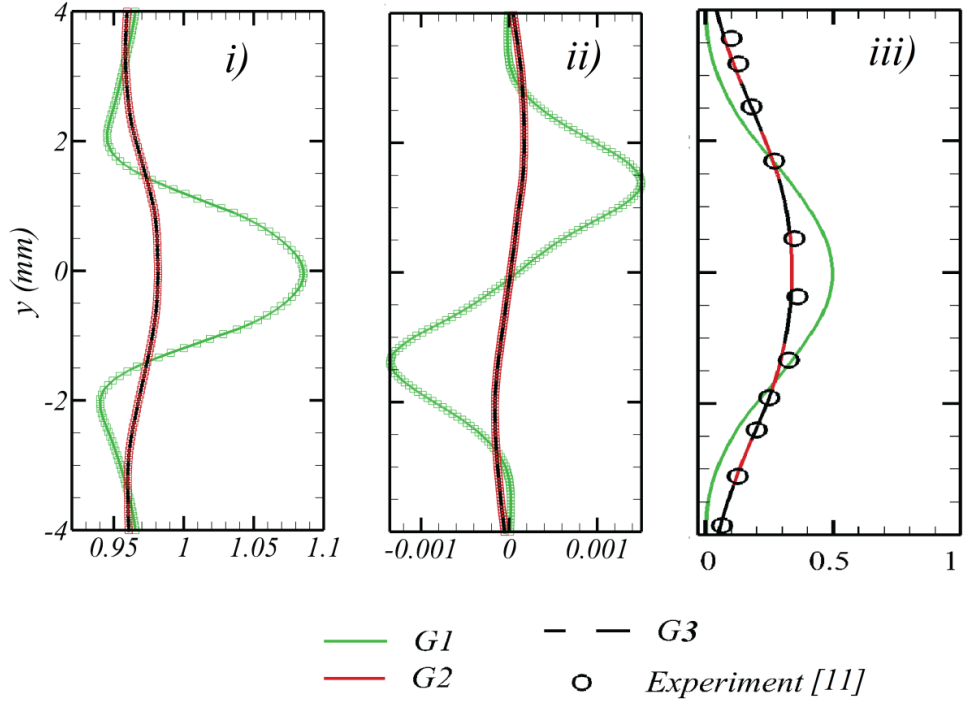

Figure 2: *i)* Normalized mean streamwise velocity ($u/u_j$) *ii)* Normalized Reynolds Stress ($\sigma_{u'v'}/u^2_{\infty,j}$) and *iii)* Mean hydrogen mole fraction for all the grids at $x/D_j = 384$ for planar jet in *TS* case

Figure 4 (b) presents the mixing efficiency defined as that fraction of least available reactant that can react if the mixture is brought to chemical equilibrium and is calculated as proposed by Mao[34]

$$\eta_m = \frac{\int Y_f \rho u \, dA}{\int Y \rho u \, dA}$$

where, (8)

$$Y_f = \begin{cases} Y, & Y \leq Y_s \\ Y_s(1-Y)/(1-Y_s), & Y > Y_s \end{cases}$$

where $Y$ represents the mass fraction of the reactant (hydrogen in the present case), $Y_f$ the mass fraction of reactant that can react and $Y_s = 0.0292$ is the stoichiometric mass fraction of hydrogen. The integration is performed on cross section area perpendicular to the direction of flow.

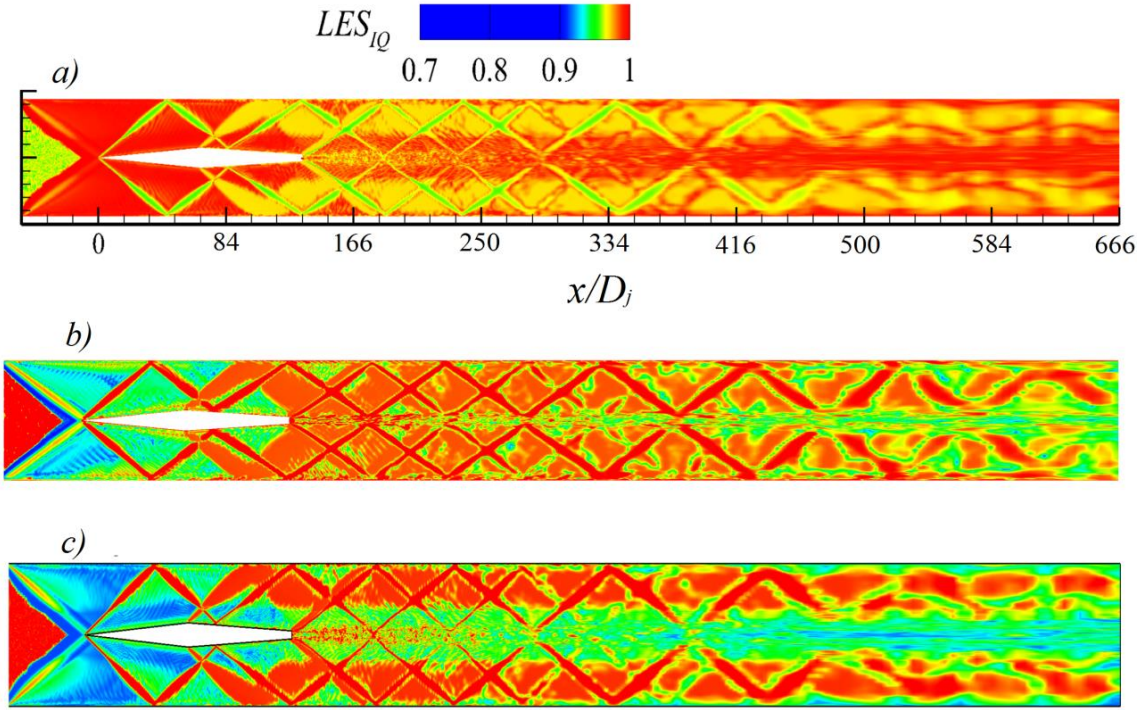

Figure 3: LES quality index for a) $X_2(2D_j)$ b) $X_3(3D_j)$ and c) $X_5(5D_j)$ configurations to demonstrate the grid resolution for all the cases: the index is well above 95 % in the region of interest

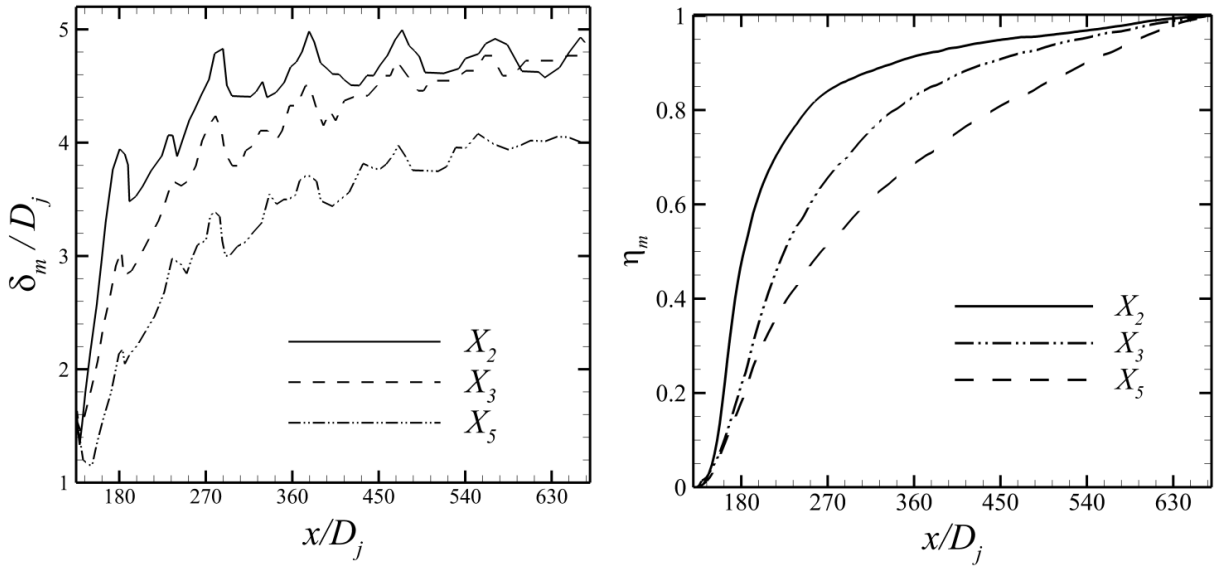

Figure 4: Spatial variation of a) normalized mixing layer thickness and b) mixing efficiency [$X_2=(2D_j)$, $X_3=(3D_j)$, $X_5=(5D_j)$].

The differences arising in the mixing layer thickness at the downstream distance is clearly manifested in the mixing efficiency. It is worth mentioning that $X_2(2D_j)$ configuration achieves approximately *85%* efficiency at around $x/D_j = 266$. However, the trend for $X_3(3D_j)$ & $X_5(2D_j)$ cases is in sharp contrast to that of the $X_2(2D_j)$ case, although at the end of channel *100%* efficiency is achieved but the near field mixing behaviour is severely affected. In case of $X_2(2D_j)$ separation, the steep rise in mixing efficiency and hence mixing is visible whereas for other configurations the mixing process in near field seems to be taking place gradually. It is well established fact that the low frequencies structures are mainly responsible for the mixing process. In the present scenario it is expected that due to the large velocity difference, the two streams will experience strong gradient at the point of separation and will tend to roll into *K-H* type vortices (or spanwise rollers). These vortices while propagating downstream facilitate the exchange of mass and momentum across the shear layer (detailed discussion will follow in the later section). From this result, one fact is established that even for a given convective Mach number (supersonic) slight variation in jet spacing can significantly alter the mixing behavior.

Having identified astonishing difference in the mixing behavior for all three configurations it is worthwhile to investigate the phenomenon responsible for such behavior. Figure 5 depicts the magnitude of instantaneous density gradient for all the three cases. The contour plot sheds light on the complexity of the flow physics, thereby revealing different flow features. At the end of the strut, expansion fan is originated and after reflection interacts with the reflected shock system originated at the strut tip. Various other interactions namely *shock/shock (SSI)*, *shock/shear layer (SSLI)* and *shock boundary layer interactions (SWBLI)* are unfurled. Among these multiple complex interactions, *SSLI* seems to play a vital role in shaping the mixing behavior whereas *SWBLI* has the least effect. The *SSLI* leads to the production of vorticity due to the non-alignment of the pressure and density gradient popularly known as baroclinic torque. This particular mechanism plays a very crucial role, especially in the supersonic cases dominated by the shock. The axial vorticity generated across the shear layer results in the lifting of the vortices and amplification of the strain rate thus enhancing the mixing.

The Overall shock system (train) is steady in nature and is not altered by the presence of the unsteady recirculation at the strut end. Recirculation bubble which is unsteady in nature, itself breaks down being trapped between the two streams moving at a different velocity. The smaller vortical structures generated during this process also adds to the quasi-periodic shedding of the vortices mimicking wake flow typical of the cylinder. From the contour plot, a sharp difference in the flow field is witnessed, especially comparing $X_2(2D_j)$ with the other two configurations. The contour plot suggests the early onset of the wake mode for a $X_2(2D_j)$ case with various spatial scales however same is not true for the $X_3(3D_j)$ & $X_5(5D_j)$. In the near injector region for $X_2(2D_j)$ cases *K-H* structures are visible because of the instability gets triggered across the shear layer due to the velocity gradient and the convection of the unsteady vortical motion through the recirculating zone. The presence of spanwise roller suggests the dominance of *two*-dimensional effects in the near field, theses quasi *two*-dimensional rollers grow in size upon convecting downstream due to the pairing and merging process of vortices. However, in the far field, they are transformed into fine scale structures lacking coherence because of the turbulence. In near field, these structures are vital to the mixing process by engulfing the fluid from the surrounding.

It appears from the plot that $X_2(2D_j)$ configuration having less jet spacing exhibits a similar type of flow field as experienced in the case of planar jet[8], whereas for $X_3(3D_j)$ & $X_5(5D_j)$ the scenario is entirely different. It can be attributed to the spanwise extent of the separated region between the injectors. It is evident that the *3* dimensional effects in the near field region inhibit the presence of spanwise rollers which may further affect the mixing characteristics. Clemens & Mungal[6] conducted an experiment by varying the convective Mach number and suggested that three dimensionality in mixing layer may alter the growth rate of entrainment.

Q-criterion (or the second invariant of velocity gradient tensor) proposed by Hunt et al.[35] is a powerful vortex identification technique. In Figure 6 Q-criterion is presented for all the cases, where for $X_2(2D_j)$ spacing shear layer instability is clearly visible in the near field. Due to this instability of the shear layer spanwise rollers are present in the near field. Early formation of these structures ensures increased mixing due to mixing/pairing and stretching phenomenon associated to the vortex motion. When comparing this case to the other two, i.e. $X_3(3D_j)$ and the $X_5(5D_j)$ noticeable difference is observed. The structures marked as *I* are spanwise rollers and the vortices marked as *II* are formed due to the shock/shock interaction. The role of *II* type vortex in mixing augmentation is not clear as they seem to convect downstream without interacting with other vortices generated across shear layer region. It can be conjectured that the $X_2(2D_j)$ spacing cases experience relatively higher velocity gradient across the shear layer near injector due to moderate three-dimensionality. However, for the higher jet spacing, the gradient might be severely affected as the flow is more relaxed in the third dimension. The major difference is noticed for the highest jet spacing case where least gradient is experienced and the shear layer mostly remains stable and lacks the formation of spanwise rollers completely. For this configuration mostly the flow is dominated by the three dimensional structures which seem to lack coherence. Once again it is established that early breakdown of jet leading to the formation of coherent structures is one of the most important parameters. Further discussion related to these realizations is subject of the following section.

Having established qualitatively that vortex break down may be a very critical parameter; it is quite obvious to look into the reason associated with break down. The various researchers conducted a study to have a better insight of the underlying physics. Delery[36] and Kalkhoran & Smart[37] conducted research to characterize the shock/vortex interaction for both normal and oblique shock waves and they concluded that some of the main parameters that trigger vortex breakdown are axial velocity deficit, shock intensity etc. Hiejima[37, 38] in their work noticed vortices interacting with the shock wave are stretched and become unstable. Hence, the need is not only to understand the breakdown process, it may be equally insightful to learn the vortex breakdown location. Pierro & Abid[39] proposed that enstrophy integrated along the plane normal to streamwise direction is a good indicator of break down point. They suggested that the local maxima peaks in the integrated enstrophy indicate the vortex breakdown location. Hiejima[40] also utilized the similar method in their investigation of hypermixer strut combustion to locate these key points in the domain. The enstrophy is integrated as,

$$\varepsilon = \int_{\Omega} \frac{1}{2} \omega_i \omega_i \, d\Omega \qquad (9)$$

$$\varepsilon_{y+z} = \varepsilon - \int_{\Omega} \frac{1}{2} \omega_x^2 \, d\Omega \qquad (10)$$

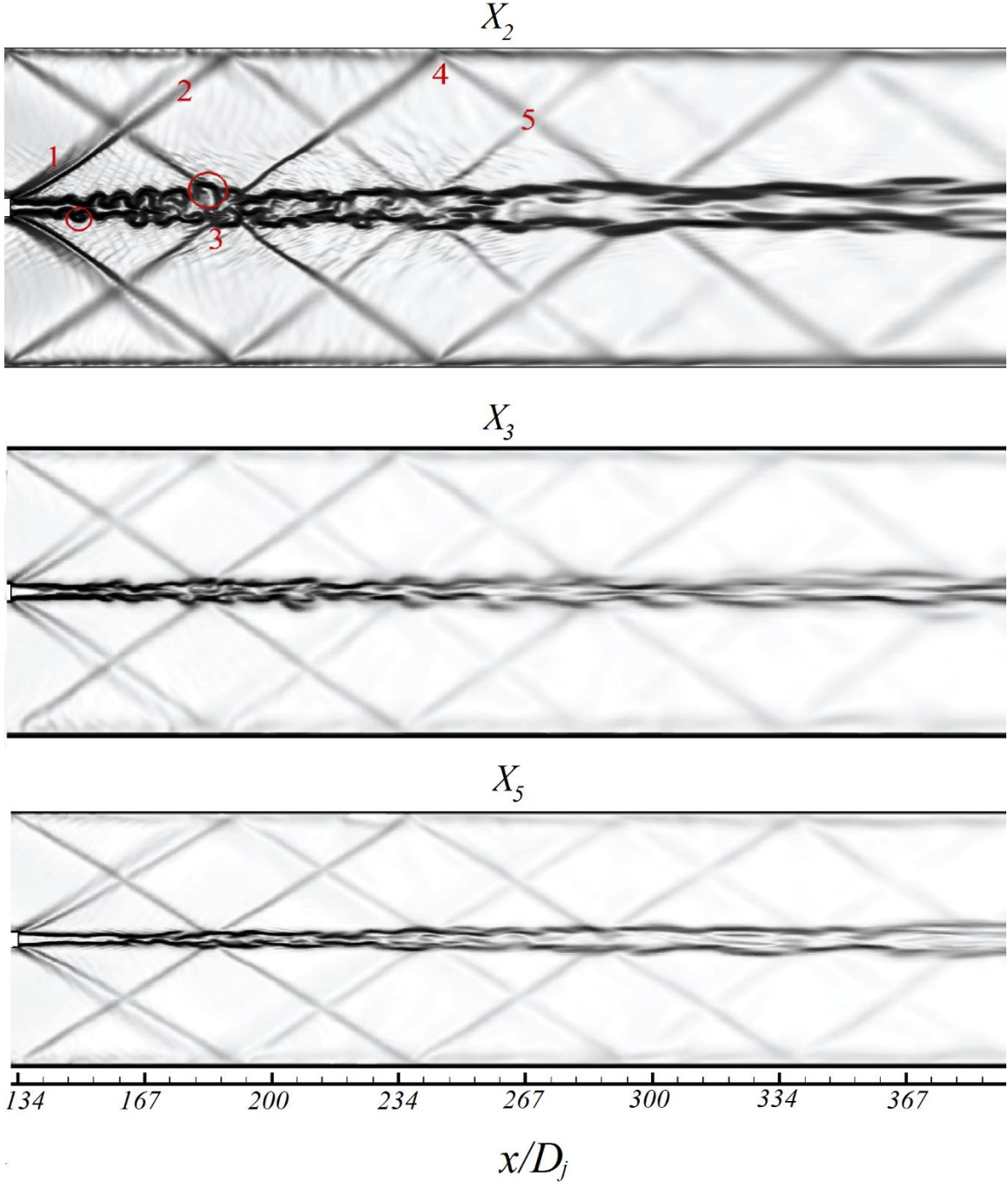

Figure 5: Instantaneous density gradient distribution past injection point for all three cases along $z/D_j=0$ plane. (1 – Expansion fan, 2 – Reattachment shock, 3 – Shock/ Shear layer interaction (*SSLI*), 4 – Shock/Boundary layer interaction (*SWBLI*), 5 – Shock/Shock Interaction and the red circle shows the *K-H* instability and growth of vortices) [$X_2=(2D_j)$, $X_3=(3D_j)$, $X_5=(5D_j)$].

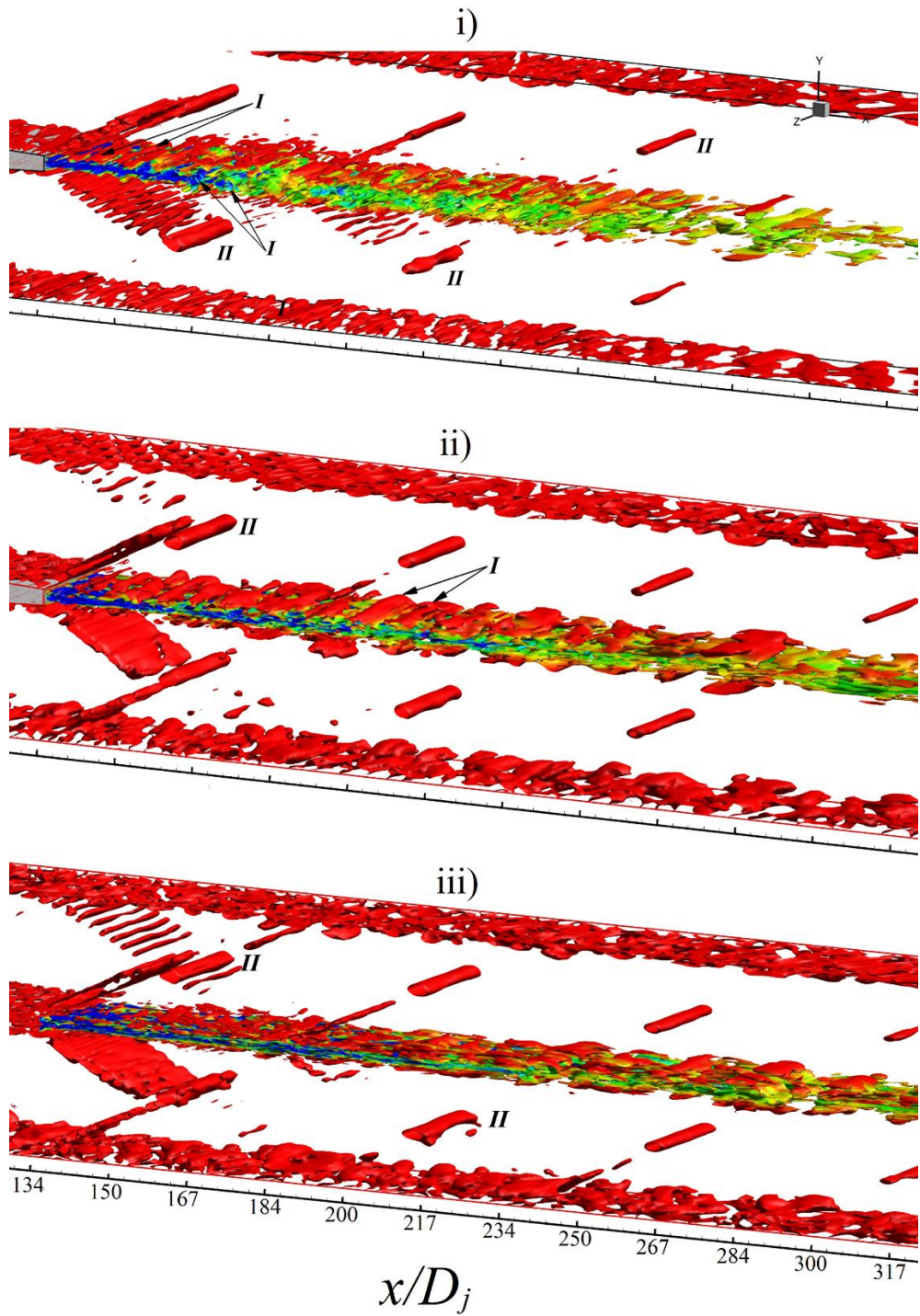

Figure 6: Q- criterion iso-surface coloured with the instantaneous hydrogen mole fraction for i) $X_2(2D_j)$ ii) $X_3(3D_j)$ and iii) $X_5(5D_j)$
(Blue color indicates the highest vale and red the least)

The integrated enstrophy as per Eq. (10) is normalized by dividing it by the enstrophy at the jet exit (for respective cases), i.e. at $x/D_j \approx 134$ for all the configurations. Looking at the enstrophy trend it is clearly conveyed that all the configurations have local maxima peaks however at different locations. The consistency between the vorticity contour (Figure 6) and spatial enstrophy variation (Figure 7) is well established. For the $X_2(2D_j)$ case, vortex breakdown occurs very close to injector but is shifted downstream for other two configurations. This clearly confirms and establishes the fact that an early onset of vortex breakdown is a very useful parameter to have better mixedness in the near field region which is vital to the efficient combustion.

Once again if this observation is combined with that of Figure 4, it is evident why $X_2(2D_j)$ configurations have higher spreading which is manifested in the mixing.

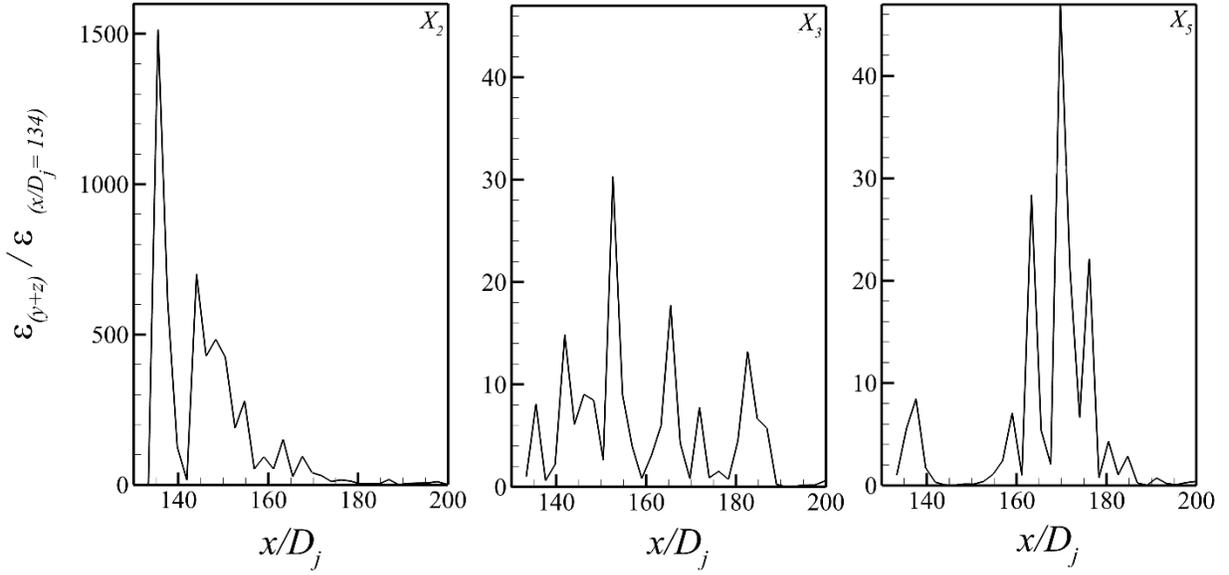

Figure 7: Axial variation of normalized enstrophy along the jet centerline [$X_2=(2D_j)$, $X_3=(3D_j)$, $X_5=(5D_j)$].

To further quantify that the intensity of turbulence generated and shear strength ($R=u_2/u_1$) in the near field is responsible for rapid mixing, Figure 8(a) presents the velocity gradient plot at $x/D_j \approx 138$ for all three configurations. The noteworthy difference in the magnitude is apparent as it suggests that in the near field region the shear layer for the $X_2(2D_j)$ case is subjected to higher shear than any other case. One possible reason is that the flow is more relaxed in the spanwise direction for higher spacing configurations, the effect of which is manifested in the velocity gradient plot. In a similar fashion, Figure 8(b) shows the normalized Reynolds stress component at $x/D_j \approx 138$, higher values are registered for $X_2(2D_j)$ case along the shear layer. Figures 8(c) & 8(d) represent streamwise turbulent fluctuation at $x/D_j \approx 138$ and along the jet centerline which clearly suggests higher turbulence intensity for the $X_2(2D_j)$ case. The qualitative and quantitative observation made so far confirms that the shear strength, turbulent intensity, and vortex break down location are some of the most critical parameters affecting the mixing behaviour in realistic supersonic injectors. Hence it becomes imperative to consider these parameters strongly while designing the Scramjet combustor.

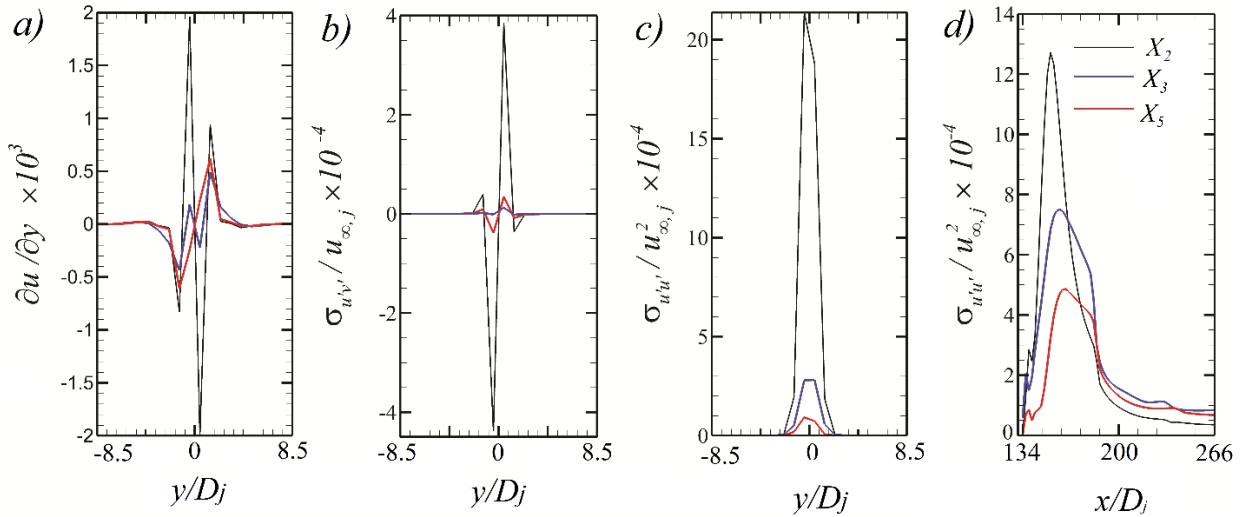

Figure 8: *a)* Transverse velocity gradient, *b)* Normalized Reynolds stress component, *c)* Normalized streamwise turbulent intensity (first 3 plots correspond to $x/D_j=138$) and *d)* Normalized streamwise turbulent intensity along the jet centerline (*subscript $\infty$ & j correspond to freestream and jet respectively*) [$X_2=(2D_j)$, $X_3=(3D_j)$, $X_5=(5D_j)$].

**B. Effect of Strut Shape**

From the first part, it turns out that even from a given high $M_c$ for moderate spanwise spacing significant difference in the mixing characteristics is noticed. As noted in the previous section that irrespective of $M_c$, higher velocity gradient (or shear strength) close to the injector is mainly responsible for the increased diffusion of $X_2$-$TS(2D_j)$ configuration. Higher velocity gradient leads to the generation of well-known $K$-$H$ vortices which, in turn, leads to the production of turbulence along the shear layer leading to enhanced turbulent diffusion. Having confirmed the dominance of three-dimensional effect in the inhibition of mixing especially for higher spanwise separation, the strut is modified such that the jet is injected past the straight contour as opposed to earlier tapered contour (Fig. 1). This modification would yield stronger recirculation region and hence stronger reattachment shock; it would be noteworthy to analyze the formation and evolution of vortical structures due to this geometrical altercation and its overall effect on the mixing behavior.

Figure 9 depicts the streamwise variations of normalized mixing layer thickness and mixing efficiency for both $X_2$-$SS(2D_j)$ and $X_2$-$TS(2D_j)$ arrangements. One can witness the exceptional difference between the two cases. Such striking variation suggests a major difference in the flow physics. Upon inspecting the profiles in Figure 9 remarkable difference in the trend is unveiled, both mixing layer thickness and mixing efficiency show significant improvement for the $X_2$-$SS(2D_j)$ configuration when compared to $X_2$-$TS(2D_j)$ configuration. The streamwise profiles of both the parameters unveil remarkable difference when compared to the $TS$ arrangement. However, for the $TS$ case, the growth is not that steep, and this noticeable difference in the growth rate of two mixing layers can be attributed to the formation of strong reattachment shock. Also, unsteadiness of recirculation region can be attributed to this observation as these factors are discussed at a later stage. Figure 9(b) presents mixing efficiency along the jet centerline. It is noteworthy that for $X_2$-$SS(2D_j)$ at around $x/D_j \approx 250$ the mixing efficiency attains an impressive value of about 95 % which is much higher compared to $X_2$-$TS(2D_j)$ case. So from these two plots, it appears that for the given $M_c$, wedge angle and spanwise separation between the jets, $SS$ arrangement performs excellently in the near-field as well as in the far-field region. The enstrophy integrated perpendicular to the streamwise plane is normalized by the enstrophy at the jet exit and is exhibited in Figure 9(c). The enstrophy trend further highlights that the jet break down occurs at approximately at $x/D_j \approx 138$ which leads to the formation of local vortices. This brings to the question why there is such a noticeable difference due to the variation of the base region thickness (lip thickness) in mixing characteristics. It strongly suggests that a minor alteration in lip thickness modifies the turbulent field and also promotes the formation of large scale structures.

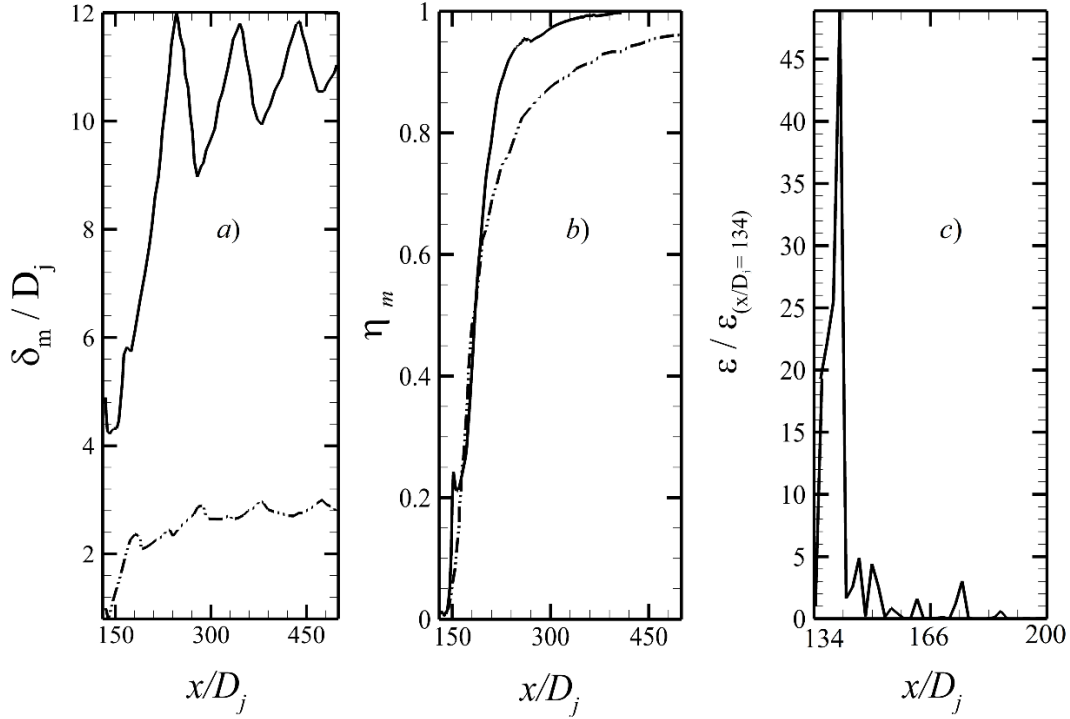

Figure 9: Spatial variation of i) normalized mixing layer thickness ii) mixing efficiency and ii) normalized enstrophy, solid and dotted line represents the $X_2$-$SS(2D_j)$ and $X_2$-$TS(2D_j)$ respectively

It is quite necessary to get an insight of turbulence statistics for better understanding of mixing behavior and role of turbulence in the mixing process. From Figure 10(a) one can see that though the velocity gradient is of the comparable magnitude at $x/D_j \approx 138$ for *TS & SS* arrangements yet huge difference has been witnessed in the near field mixing activities. Figures 10(b & c) present normalized Reynolds stress and streamwise turbulent intensity at a similar location, while a noteworthy difference is unfurled in the shape and magnitude between two configurations. The turbulent intensity corresponds typically of the wake mode for the *SS* configuration clearly depicting the role of shock induced instability leading to wake formation. The low frequency vortices generated due to the *shock/shear layer interaction* trigger this vortex break down which, while rotating along the spanwise axis, pull fluid from the outer region to the inner region thus facilitating mixing. The turbulent intensity profile plotted along the jet centerline is shown in Figure 10 (d), which again reemphasizes that most of the turbulent production takes place in the *SSLI* region due to the non-alignment of the pressure and density gradient.

Having established the role of turbulence in the mixing process due to the formation of the structures along the shear layer, the detailed understanding towards the evolution and structure of low frequency vortices becomes necessary. Substantial knowledge of turbulent flow can be acquired by merely studying the spatial coherence in the flow. It is with the detailed description of the coherent structures one can establish the order in highly chaotic systems. More detailed discussion on coherent structures can be obtained from the work of Hussain[41] & Fiedler[42]. Moving on, the instantaneous density gradient contours at various normalized time ($t = T*(u_j / D_j)$) instants are presented in Figure 11. Various hydrodynamic features are revealed such as *SWBLI, SSLI, SSI* and eddy shocklets etc. The approaching boundary layer is separated in the base region due to finite lip thickness and is unsteady in nature; where a sharp contrast can be observed between the two flow field, i.e. $X_2$-$SS(2D_j)$ and $X_2$-$TS(2D_j)$. The shear layer formed at the tip of the base region is unstable and upon the *shock/shear interaction* further instability is induced which leads to the jet breakdown and formation of low frequency spanwise rollers mimicking K-H type vortices. It is clear from the counter plot that *K-H* instability is triggered at the point of interaction (*SSLI*) and beyond this point wake mode is found to be visible. Although the presence of shock wave in the system results in pressure losses but it is unavoidable in the Scramjet propulsion system. As a matter of fact, the studies by Lu & Wu[43], Drummond and Mukunda[44] and Buttsworth et al.[12] reported incredible increment in mixing. *Shock/ shear interaction* can additionally lead to substantial increase in the turbulent intensities and extraction of energy from the mean flow. Although the effect is local and is limited to a few jet heights downstream of the interaction point yet the production of streamwise vorticity can induce sharp growth in the mixing layer. It is also pointed out by Lu & Wu[43] that shock waves must be employed at the earliest possible upstream location to be more effective. Genin & Menon[45] performed a large-eddy simulation of *shock/ shear layer interaction* and reported that although in the self-similar region the effect is not prominent but in the non-self-similar region (near field) favorable impact is noticed. This impingement is reflected in the amplified transverse velocity fluctuation and statistics of mean shear stresses.

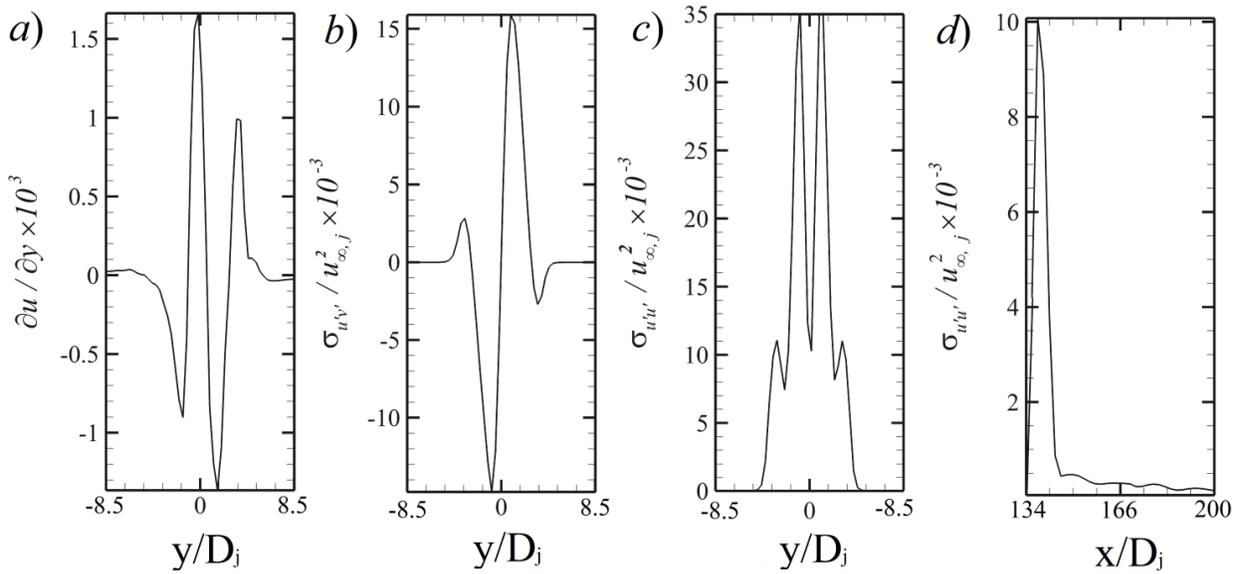

Figure 10: $X_2$-$SS(2D_j)$ case: *a)* Transverse velocity gradient, *b)* Normalized Reynolds stress component, *c)* Normalized streamwise turbulent intensity (plots a) - c) correspond to $x/D_j=138$ and *d)* Normalized streamwise turbulent intensity along the jet centerline (*subscript $\infty$ & j correspond to freestream and jet respectively*)

Again visiting Figure 11 one can see that the flow field is loaded with many complex interactions, the presence of eddy shocklets are known to modify the turbulence locally which can be verified from the contour plot. The shock wave appears to be a corrugated and steady lot of acoustic wave can be seen trapped within the first shock cell. At the interaction location initially, the shear layer tends to roll in vortex sheet of relatively smaller size, however, few jet diameters from that point, the vortical structures appear to be more organized, quasi-periodic and coherent. While convecting downstream due to the interaction, these vortical structures appear to be enlarged which is due to the pairing and merging of the vortical motion. At around $x/D_j \approx 266$, the coherence seems to be vanishing which is due to the second crossing shock wave which leads to the stretching along the streamwise direction into larger filaments. The rollers undergoing tremendous stretching at this point of break down process and assume new formation as streamwise vortices spinning along the streamwise axis. Apart from the second interaction, the highly strained braid region in between successive rollers stretches them out while convecting downstream.

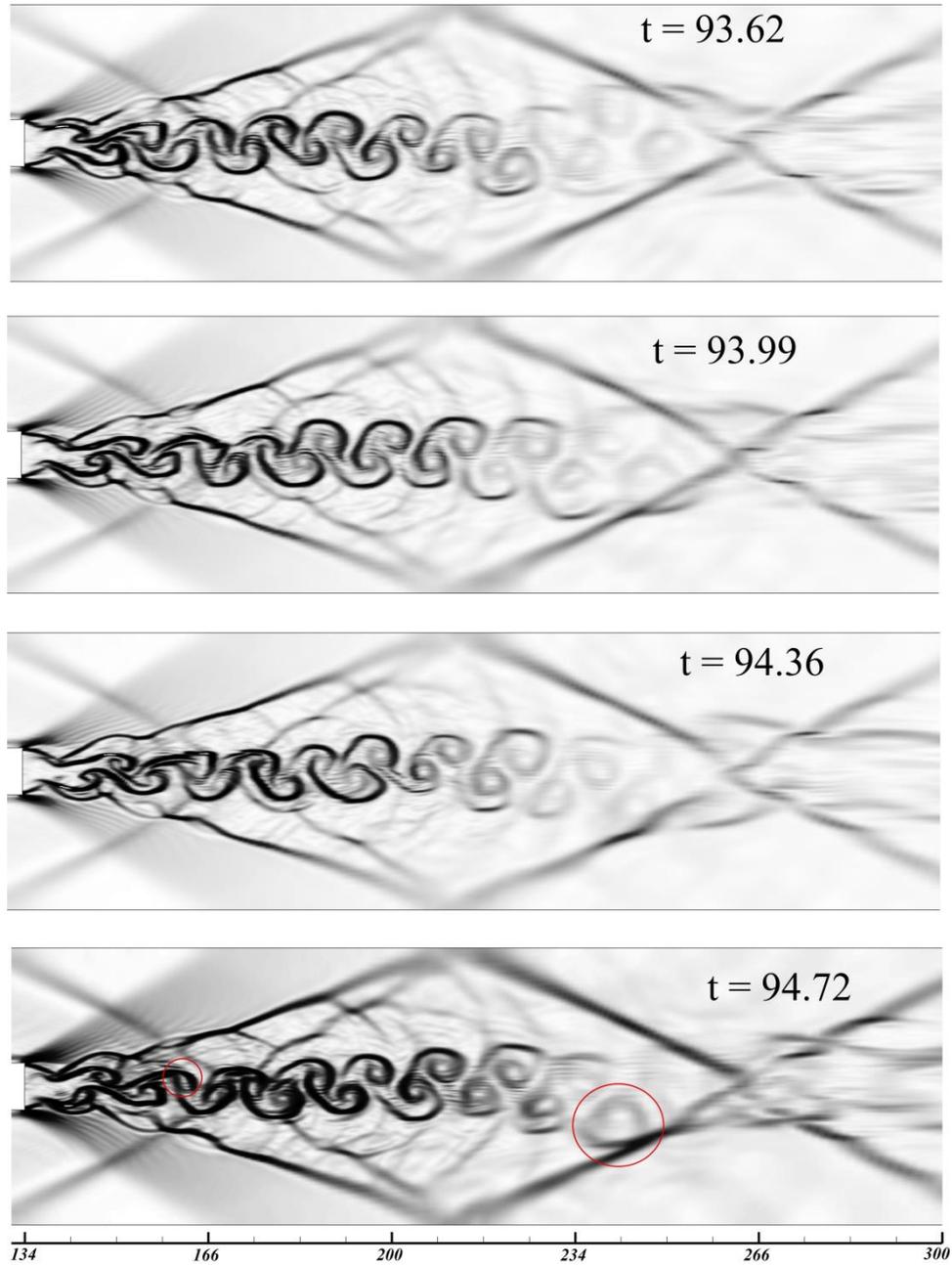

Figure 11: $X_2$-$SS(2D_j)$ case: Instantaneous density gradient contour depicting the evolution of large scale structures at different normalized time instants [ $t = T*(u_j/D_j)$ ]

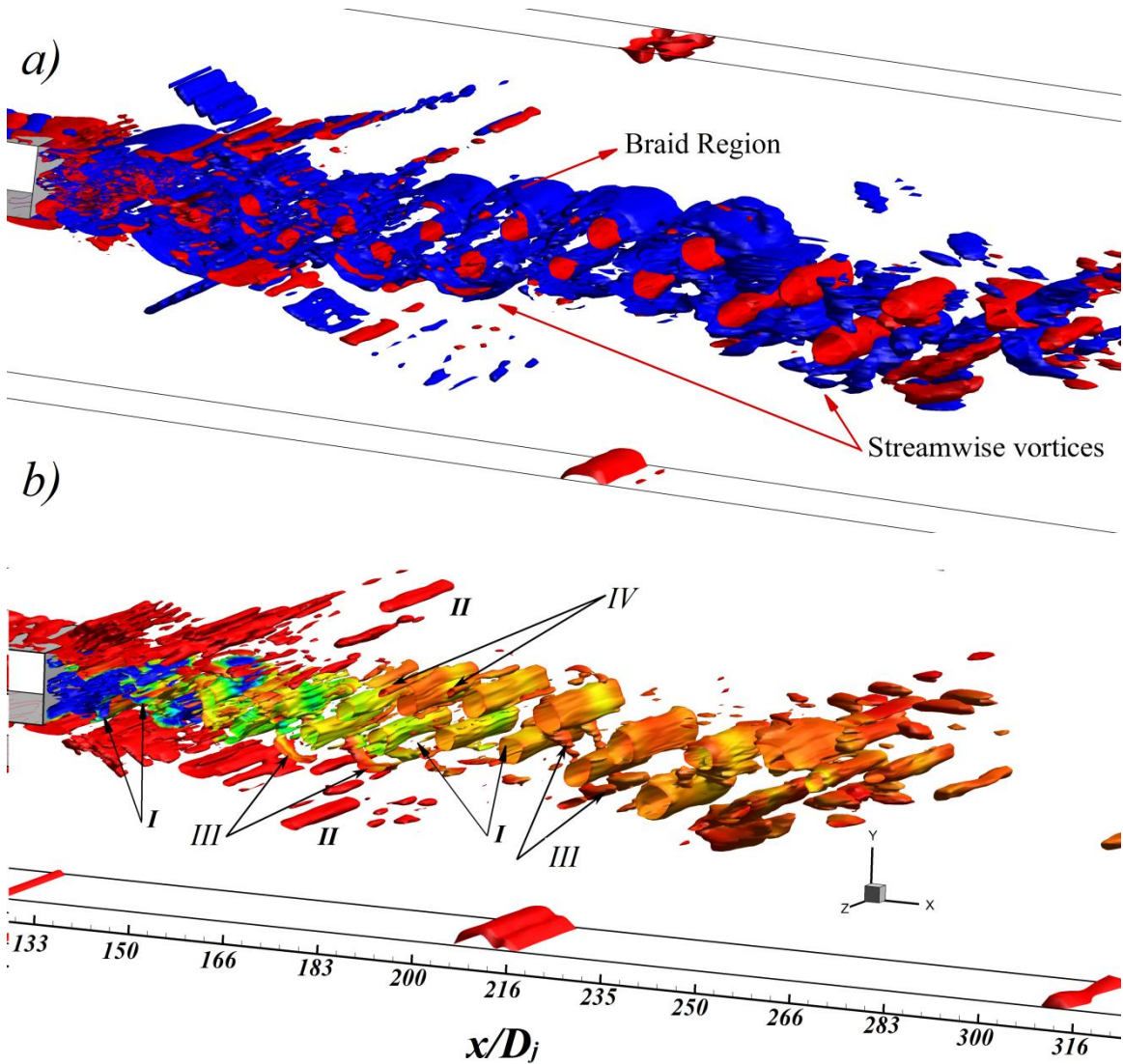

Figure 12: *X$_2$-SS(2D$_j$)* case: Q-criterion iso surface depicting various flow features and vortical structures. (a) Blue and Red color corresponds to negative and positive values and (b) is coloured with the instantaneous hydrogen mole fraction where blue indicates the highest value

In Figure 12(a) iso-surface of the second invariant of velocity tensor is plotted with both positive and negative values to reveal some of the interesting flow features. The positive values of Q correspond to the regions dominated by rotation and the negative region depicts strained region. The blue region along the shear layer and in-between the rollers indicates the highly strained region (*braid*) and region in red are spanwise rollers of high vorticity. Streamwise vortices (or *ribs*) are also present mostly in the second *shock/shock interaction* region. These rib vortices undergo stretching due to rotation of the spanwise roller and while doing so these structures pull mass from the surrounding which is eventually engulfed by the larger vortices (rollers). In Figure 12(b) vortical motion can be seen more clearly where the vortices labelled *I & II* are spanwise rollers and generated due to the *shock/shock* interaction respectively. However, the vortices generated due to the *SSI* do not seem to play any vital role in the mixing process. Apart from these quasi two-dimensional spanwise structures, vortices along the streamwise direction are also present in the near field labeled as type *III* structures. These are the rib vortices which have rotation along the streamwise direction and engulf flow from the outer region toward the inner region which is then carried by the type *II* structures facilitating mixing process. It is worth noticing that around $x/D_j \approx 216$ the ribs vortices are tremendously elongated along the streamwise direction. This is due to the extreme strain experienced by ribs due to the presence of the spanwise rollers. Finally, the type *IV* vortical structures are similar to the type *I* vortices, however, it appears that these are small scale structures are at super harmonic to that of type *I*.

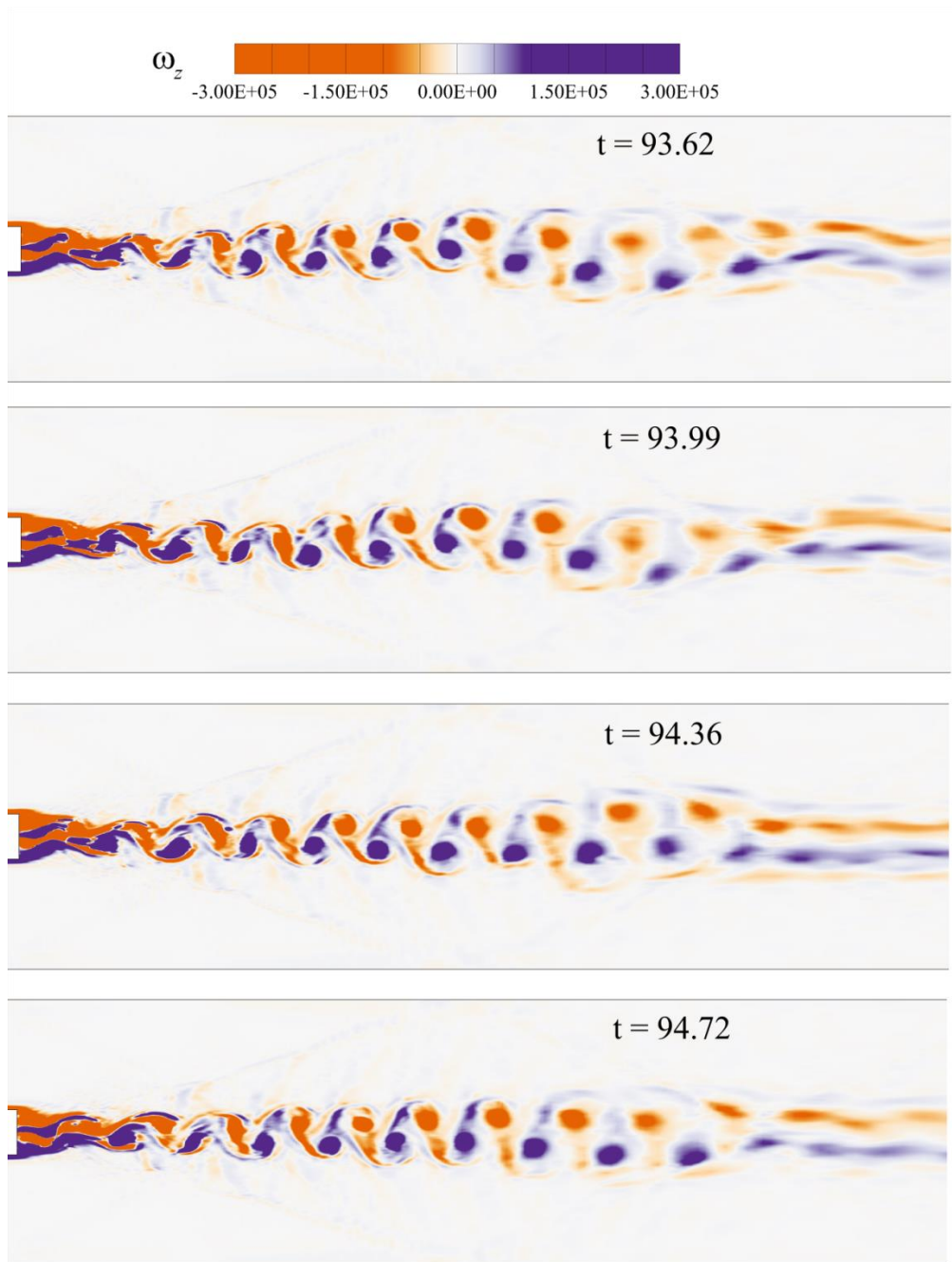

Figure 13: $X_2$-$SS(2D_j)$ case: Instantaneous spanwise vorticity at different time instants [ $t = T*(u_j/D_j)$ ] depicting the transition wake mode

The out of a plane component of vorticity is shown in Figure 13, which clearly indicates the wake mode immediately close to the injector region. The jet transitions into the wake mode dominated by *Von-Karman* type Vortex Street within five jet diameters from the jet exit. The velocity induced due to the spinning of the two consecutive rollers leads to the strained region between them. This braid region is the backbone of the evolution of the streamwise structures (or ribs) which are counter-rotating and co-exist with the rollers over brief channel length. These ribs are equally important members as they advect the flow from the outer region (co-flow) to the mixing region. However, these ribs play a more crucial role in the downstream region especially close to the second *SSI*. Riggins & Vitt[46] in their study on three-dimensional supersonic combustor also reported the importance of the vortex in the mixing process. Overall, it is understood that near and far field mixing dynamics are completely different. In the near filed, rollers are behind the entrainment process due to flow being mostly quasi-two-dimensional; whereas in far field three-dimensionality is set in and streamwise vortices, i.e. ribs play a central role in the mixing process. This concludes the first part of the present study. Now, since it is established that $X_2$-$SS(2D_j)$ configuration is more favorable for the better mixing, in the last section modal decomposition is performed for this particular case only. It is expected that the modal decomposition will further strengthen the understanding of the physics by separating the different modes of flow which are dominant.

**C. Modal Decomposition of Straight Strut [$X_2$-$SS(2D_j)$]**

In the final section, the results of modal decomposition are reported; two popular and powerful methodologies that have been adapted herein namely proper orthogonal decomposition (POD) and dynamics mode decomposition (DMD) are discussed. First, the spatial coherence (POD) is discussed and then in the following section, the modes corresponding to the pure frequency, i.e. temporal coherence (DMD) is presented. To compute the spatial coherence the modes are arranged on the basis of decreasing Eigen value (or energy distribution) such that most energetic modes are revealed first. In this way usually, first few modes are usually sufficient to get insight into the various modes of the turbulent flow field. To keep the discussion concise, formal mathematical treatment has been omitted (the interested reader can refer to the work of Soni & De[8] where the formulation employed herein is presented in detail).

It is noteworthy to mention that a large enough data set is required to track coherence in the flow. In the present investigation $N=300$ snapshots are gathered, while the data has been generated for modal decomposition with the time interval ($\Delta t$) of $1\times10^{-6}$ where the choice of $\Delta t$ comes from the Nyquist criteria. It is evident from the literature that the successful decomposition relies heavily on the temporal separation and sampling frequency ($F_s$). One of the strong requirements is to sample the data at least at twice the frequency that need to be resolved. The choice of 100 kHz for the present study comes from the fast Fourier transform of the velocity signal probed at various locations in the jet near field along the shear layer region. We first start by demonstrating the snapshot independence for the POD methodology by performing the decomposition through a different number of snapshots and the outcome of which is presented in Figure 14. The energy distribution and cumulative energy content across modes are put forward for the three different values of snapshots. The value of $N$ varies as *100, 200* and *300* from the plot one can see that first few modes capture approximately *99%* of whole energy hence the inclusion of further modes may not bring any significant change in the POD modes. Similarly, cumulative energy content clearly indicates that first five modes account for 99 % energy which implies that the remaining 1% is distributed across the other modes. Although the minor difference in the trend is seen for higher mode numbers, it would be insignificant to bother about them as they are representative of turbulence i.e. very high frequency fine scale structures which is of least or no interest in the context of the present study. Therefore, the modal analysis is reported for the $N=100$ snapshots only along the $z/D_j=0$ plane. Also, it is worth noticing that the decomposition has been reported on the original computational grid, means the computational grid is not interpolated to any other grid for modal decomposition.

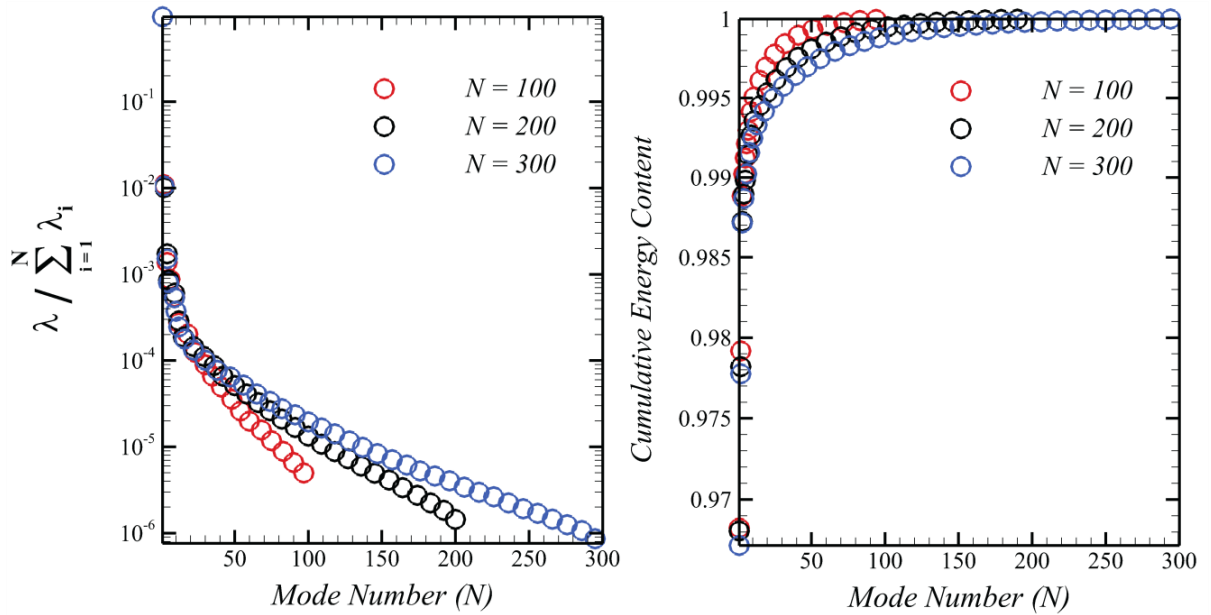

Figure 14: Demonstration of snapshot independence for the velocity (energy) based POD with three different sets of snapshots through relative Eigen vale and cumulative content

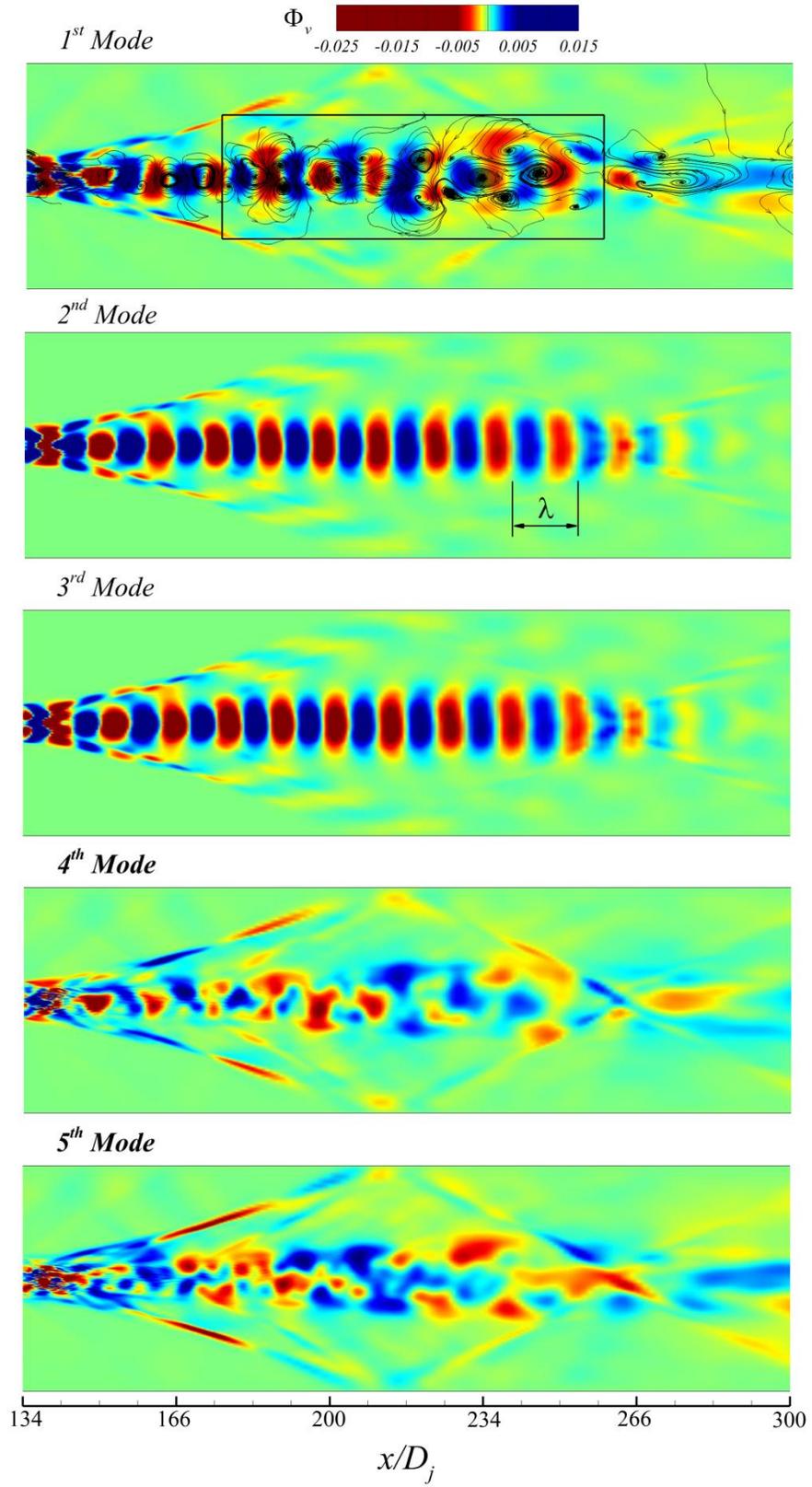

Figure 15: Spatial structures of 1st five energetic modes of the energy based decomposition. (Contour is colored with the transverse velocity)

Figure 15 exhibits the first 5 POD modes, where the contour is colored with the transverse velocity component and streamlines depict the local structures. Along the reattachment shock, spatial coherence is revealed which is consistent with the observation of Q criterion in Figure 12. Mode 1 exhibits multiple frequency structures being the most energetic mode, the symmetry in the structures actually due to the anti-symmetry in the actual field, i.e. a transverse velocity. Majority of the structures represent the wake flow with small scale vortices present along the shear region. However, the major noticeable coherence is embedded within the region $x/D_j \approx 134$ and $x/D_j \approx 266$, i.e. reattachment shock and second *Shock/shock interaction*. In the base region due to unsteady and unstable reattachment bubble, vortical shedding from this region is added to the jet core around the reattachment shock. These alternately signed mode shape represent the travelling wave depicting one wave length, in the near field contour plot exhibits symmetry about the centerline axis which is typical of wake type flow. Large scale structures are present within the mixing region and along the shear layer boundary, within first three wave lengths, single large vortical structures are noticed. However, for the following four wave lengths the smaller vortices are observed to be detached. This region is represented by the rectangular enclosure in the contour plot. It appears that this region actually depicts different vortex phenomenon, i.e. vortex tilting and vortex pairing/merging. Finally, at further downstream, small scale vortices appear past second *SSI* which leads to the enhanced strain around this region. But since these vortices are present along the outer layer they could be significant in generating turbulence locally and hence boosting mixing locally.

The next two modes, i.e. mode 2 & 3 are phase shifted but appear in a pair; while the symmetric distribution represents the wake structures, i.e. *Von-Karman* Vortex Street. The wave length $\lambda$ remain constant at around *8 mm* till $x/D_j \approx 266$; however, the structures are stretched along the transverse direction. This suggests that the vortical pairing/merging is present in between the region corresponding to $x/D_j \approx 200$ and $x/D_j \approx 266$. Beyond this point ($x/D_j \approx 266$), the vortex stretching is witnessed under the action of strain. These two modes are at a single frequency without the high frequency fine scale vortices. If these two modes are multiplied with the temporal coefficient they will exhibit the convection of the vortical motion. These modes being more organized and bearing similar frequency appear to be representative of the most probable structures. The symmetric distribution of the vertical velocity field indicates antisymmetric distribution in the vorticity field. This suggests the anti-symmetric vortex shedding cycles whereas the anti-symmetric filed corresponds to the mirrored symmetry in the vorticity field. These modes point toward the quasi-periodic wake *Von-Karman street* which is reflected in the shear layer flapping as well. These are the most dominant structures which are basically rollers, facilitate in the near field mixing. Konstantinidis et al.[47] and Feng et al.[48] also observed similar wake modes in their study pertaining to the cylinder. Further, mode 4 & 5 are disorganized and may be representative of composite vortical motion representing various frequencies. It is possible that these modes represent streamwise ribs which are actually stretched due to the rollers.

In table III first five Eigen values for both energy and enstrophy based modal decomposition is presented. For energy based POD first five modes represent the 99% energy distribution whereas in case of enstrophy about 65 % energy is captured. This difference in the energy content is due to the fact that the enstrophy based POD is performed on the first order derivative of the fluctuating velocity field hence the basis functions that form the enstrophy POD modes would be different than the energy POD modes.

Table III. Energy composition for the 1st five modes for both energy and enstrophy based decomposition for 100 snapshots

| Mode Number (N) | Energy Based | Vorticity Based |
|---|---|---|
| 1 | 0.968 | 0.292 |
| 2 | 0.01 | 0.24 |
| 3 | 0.0096 | 0.046 |
| 4 | 0.0013 | 0.043 |
| 5 | 0.001 | 0.036 |

In Figure 16 compilation of 1st five enstrophy (vorticity) based POD modes are shown where one can identify that the first two modes appear in pair with the spatial shift. The spatial shift is representative of the advection of the vortical motion of wake flow mimicking the wave like pattern. These two modes represent the large scale spanwise rollers generated due to the rolling of shear layer undergoing high velocity gradient at the interface. This motion is typical of base region flow past bluff bodies' where modes 1 and 2 mark the basic wake instability corresponding to the *2S* mode, pretty much similar to the unforced wake with similar wave length ($\lambda$) as noticed for the energy based POD. In the present scenario, the jet undergoes natural instability and behaves very well as Karman vortices after the start of the periodic shedding. The *2S* is a symbolic terminology meaning two single pairs of vortex convecting downstream, while the term was coined by the Williamson & Roshko[49] in their study related to the investigation of the flow field in a laterally oscillating cylinder environment. The symmetrical mode shape about the axis is actually anti-symmetric velocity filed distribution; the symmetric or anti-symmetric distribution arises due to the non-linear interactions of the vortical motion in the wake region.

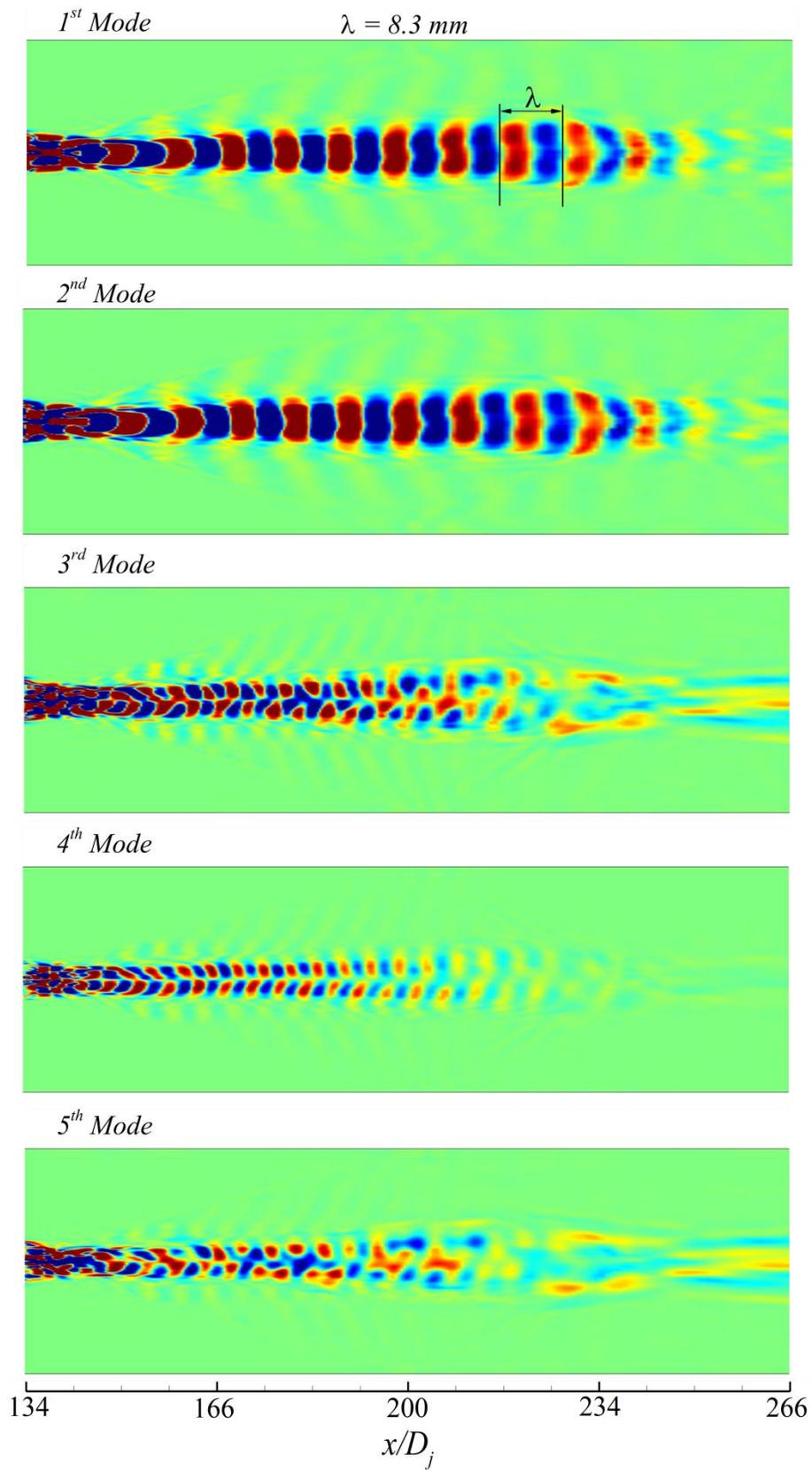

Figure 16: Spatial structures of the first 5 energetic modes of the enstrophy based decomposition. (Contour is colored with the spanwise vorticity)

Modes 3rd & 4th also appear in pair however relatively small scale vortices are observed mostly convecting along the outer edge of the shear layer. The contour of the jet region is anti-symmetric about the jet centerline and hence represents the symmetric vortex motion along the shear layer. The modal shape in 3rd & 4th mode appears in a periodic doublet with anti-symmetric distribution along the jet axis, while it is difficult to characterize these structures but one may think of them as fine scale motion at the higher harmonics. Moreover, the wave length ($\lambda$) inferred from the two streamwise neighboring structures is around half of that observed in 1st two modes. Essentially, it seems that these two modes are super harmonics of the 1st & 2nd modes. It is possible that these vortices convecting outward from the jet centerline are responsible for the growth of the mixing layer thickness. Finally, 5th mode although being somewhat similar to 3rd & 4th modes does reveal some form of spatial coherence mostly along shear layer; however, it does not offer any concrete insight into the vortical motion.

To further quantify the observation of Figures 15 & 16 fast Fourier transform of the 1st four temporal coefficients are performed and reported herein. Clarification regarding the choice of small $\Delta t$ must be clear now upon inspecting the Figure 17. Two dominant frequencies are observed from Figure 17 and upon combing this observation with that of the enstrophy modes it appears that first two modes exhibit the vortical motion oscillating at the *56.48 kHz* and the subsequent two modes, i.e. 3rd & 4th present the structures oscillating at *115.7 kHz*. Hence from the results of enstrophy modes, it can be inferred that the wake mode due to naturally occurring instability corresponds to the *2S* mode and the higher frequency which appear to be super harmonic of the fundamental frequency. Few low frequency peaks are also present especially for the 3rd and 4th mode this could be due to the fact that POD modes although spatially coherent may represent the mixed frequencies. It is possible that these low energy peaks are related to the low frequency structures possibly streamwise vortices (ribs).

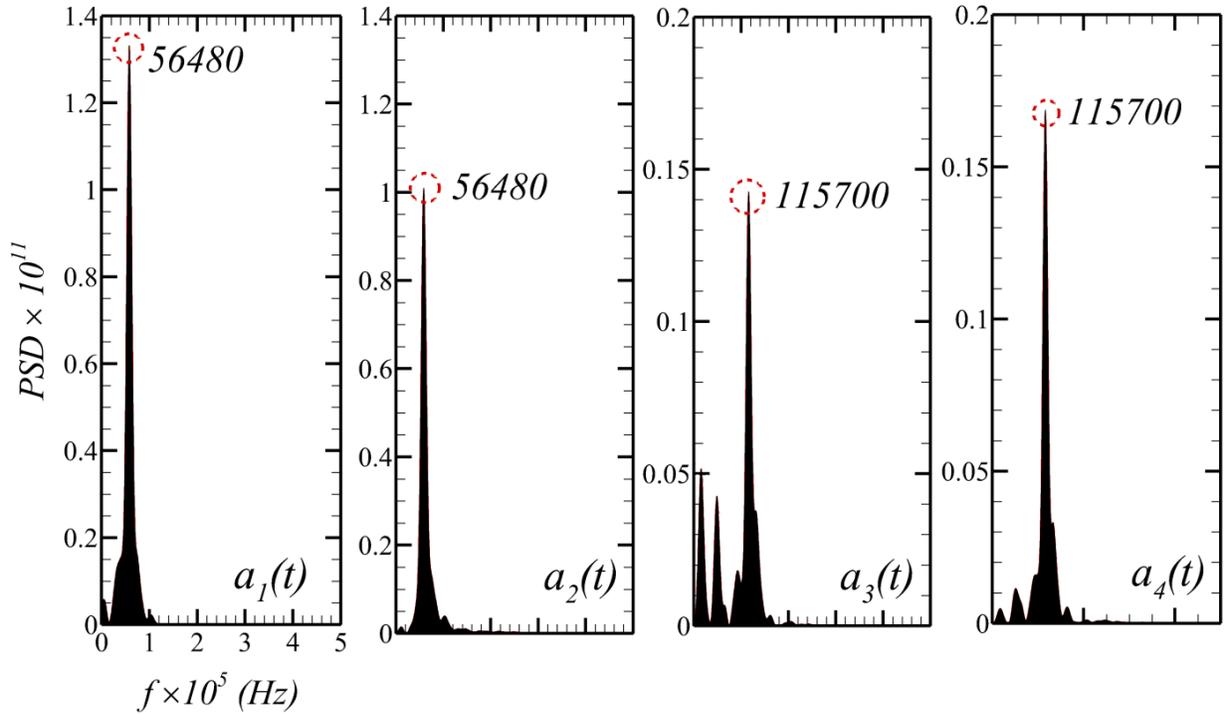

Figure 17: Fast Fourier transform of the first five temporal coefficients corresponding to enstrophy POD

Apart from vector POD which is discussed in the previous section, the scalar POD is also performed and presented in Figure 18 to demonstrate that sufficient information can still be gathered through this approach. The modes 1 & 2 exist in pair with similar wave length and frequency only with the phase shift which indicates the advection of the structures. The wave length is comparable to the 2nd & 3rd mode of the energy and 1st & 2nd of the enstrophy based decomposition. Although 3rd & 4th modes exhibit similar behavior; however, these modes have reduced wave length (approximately half) and relatively higher frequency. To be precise it can be said that vortical motion in first two modes is more dominant structures occurring at the fundamental frequency. Two regions are highlighted in Figure 18 namely *I* and *II*, as it seems that in the region *I* initial vortex stretching is taking place whereas in zone *II* the pairing and merging process is unveiled. First two mods exhibit an anti-symmetric distribution in space which signifies the presence of the *Von-Karman* street structures which is consistent with the observation of the preceding section. However, in the last two modes, the structures exhibit some local symmetry and are phase shifted which again signifies the convection of the vortices.

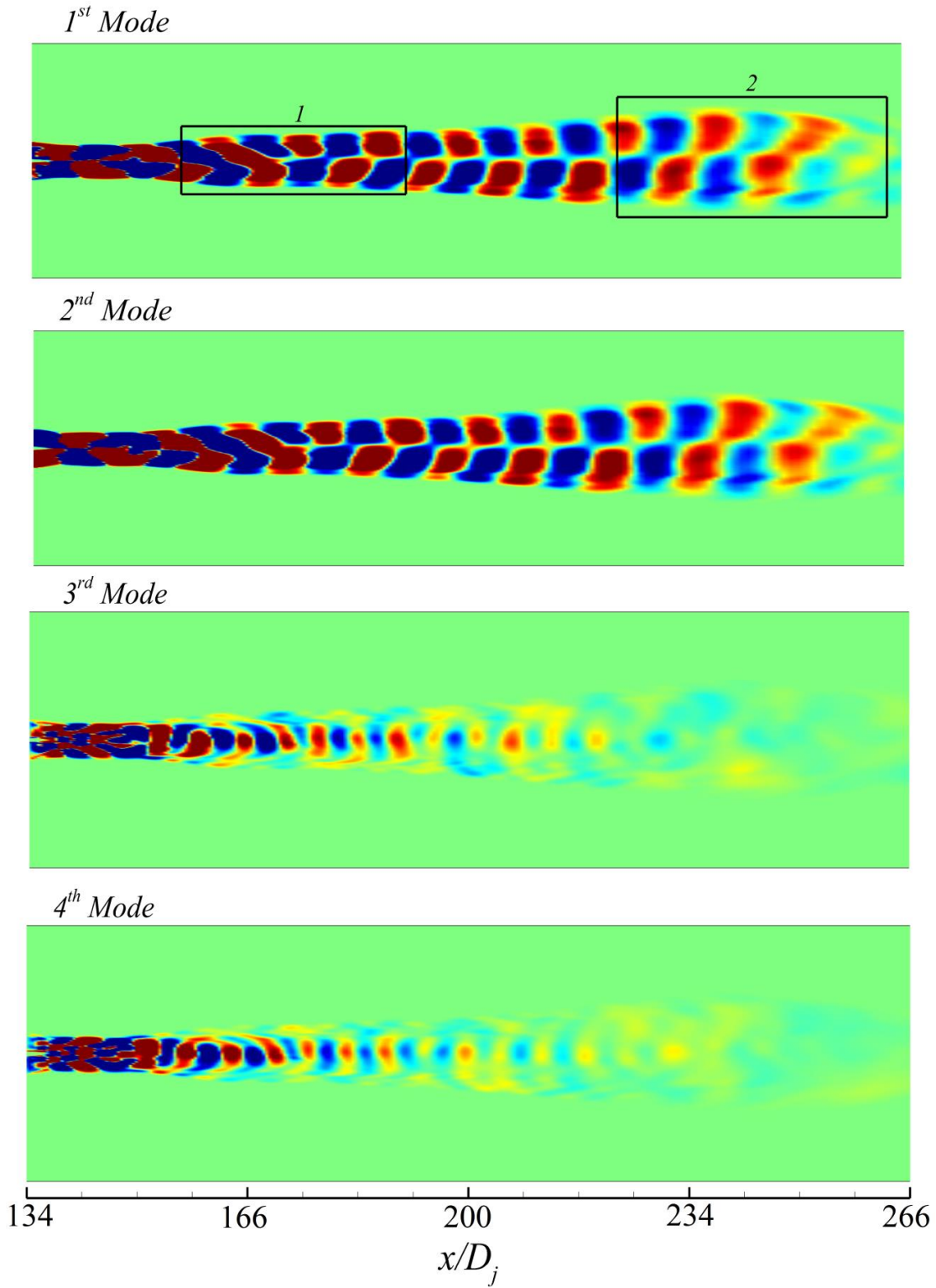

Figure 18: First four most Eigen modes for scalar POD calculated by employing only hydrogen mole fraction

The fast Fourier transform of the temporal coefficients is estimated and presented in Figure 19. One can see that similar frequencies as observed for enstrophy based decomposition are also present. Presence of frequency for first two and last two modes indicates a strong correlation between the modes. If one calculates the cross-correlation for the 1st & 2nd mode single frequency will be revealed and the same is true for the 3rd and 4th modes; however, three are some low frequencies as well. In table IV, the observed frequencies, wave length and normalized convection speed for the vortical structures present across 1st four modes (enstrophy and scalar) are presented. From the values of convection speed, it can be conjectured that the largest eddies which are basically *Von-Karman* Street convects at around 92 % of the oncoming flow and is supersonic whereas the higher mode vortices convect at higher speed. It is difficult to identify the vortical motion present in the higher modes which can be verified from the Figure 20. It is interesting to note that the first row of Lissajou's curve in Figure 20 exhibits circular trajectory which suggests the presence of strong phase relation between modes while the last row does not reveal any such information due to distorted phase portrait. The information contained in the higher modes cannot be easily observed due to lack of coherence. The circular trajectory indicates the phase shift of $\pi/2$ which implies that the two modes exist in a pair and are an indicator of the convection of vortices.

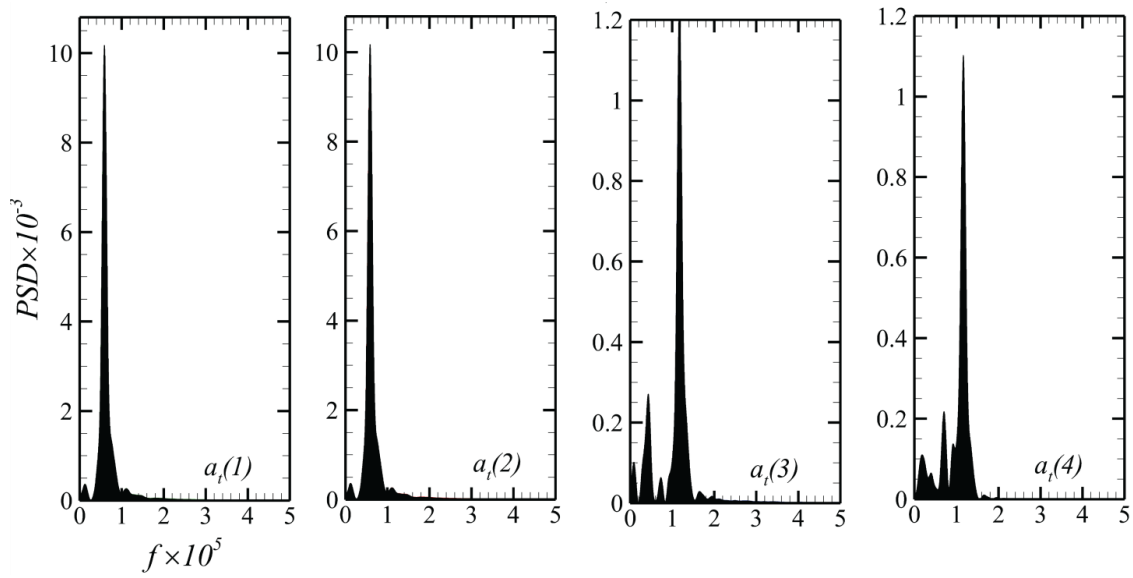

Figure 19: Fast Fourier transform of the first four temporal coefficients for hydrogen mole fraction POD

Table IV. Observed frequency, wave length and corresponding normalized convection speed for the first four modes based on the enstrophy and scalar POD.

|        | $f (Hz)$ | $\lambda$ (mm) | $c/u_\infty$ |
|--------|----------|----------------|--------------|
| Mode 1 | 56480    | 8.3            | 0.928        |
| Mode 2 | 56480    | 8.3            | 0.928        |
| Mode 3 | 115700   | 4              | 1.374        |
| Mode 4 | 115700   | 4              | 1.374        |

The proper orthogonal decomposition results discussed in the preceding section establishes the spatial orthogonality with mixed frequencies however temporal orthogonality lacks in such decomposition. Dynamic mode decomposition is temporally orthogonal and is capable of extracting the dynamic information from the flow. In DMD instead of collating modes on the basis of energy distribution as done in POD, dynamic modes are arranged on the basis of observed frequencies, in this way the Eigen values here represent the growth or decay of any instability present in the flow field. The major difference between the two approaches arises from the fact that POD modes retain multiple frequencies whereas dynamic modes are classified by the pure frequencies. For the present analysis, $N=101$ snapshots are utilized such that the resulting companion matrix has the dimension of $(N-1) \times (N-1)$. As mentioned earlier, care has been taken while acquiring the snapshots, so as to comply with the Nyquist criterion which is already demonstrated in Figures 17 & 19.

In Figure 21, the frequency spectrum of the dynamic modes is presented and the symmetric nature of the plot about the mean mode is due to the processing of the real-valued data. Various peaks are revealed however keeping up with the FFT of the temporal coefficients where only two frequencies are marked. Although two super harmonics (marked in red) are observed for the $f_1$, here only fundamental frequency and first harmonic is discussed. Some low frequencies are also present in the spectrum however for the sake of brevity those modes are not discussed.

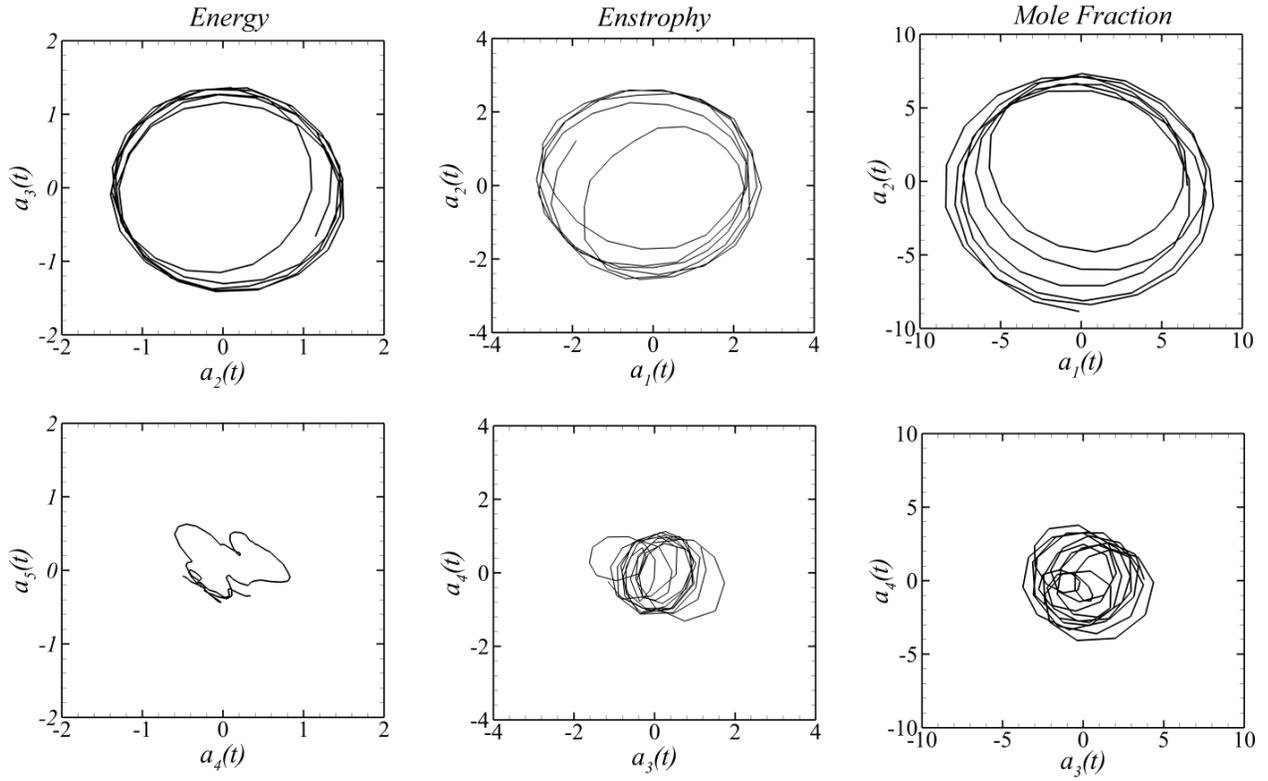

Figure 20: Phase portrait of the POD modes computed through different norms

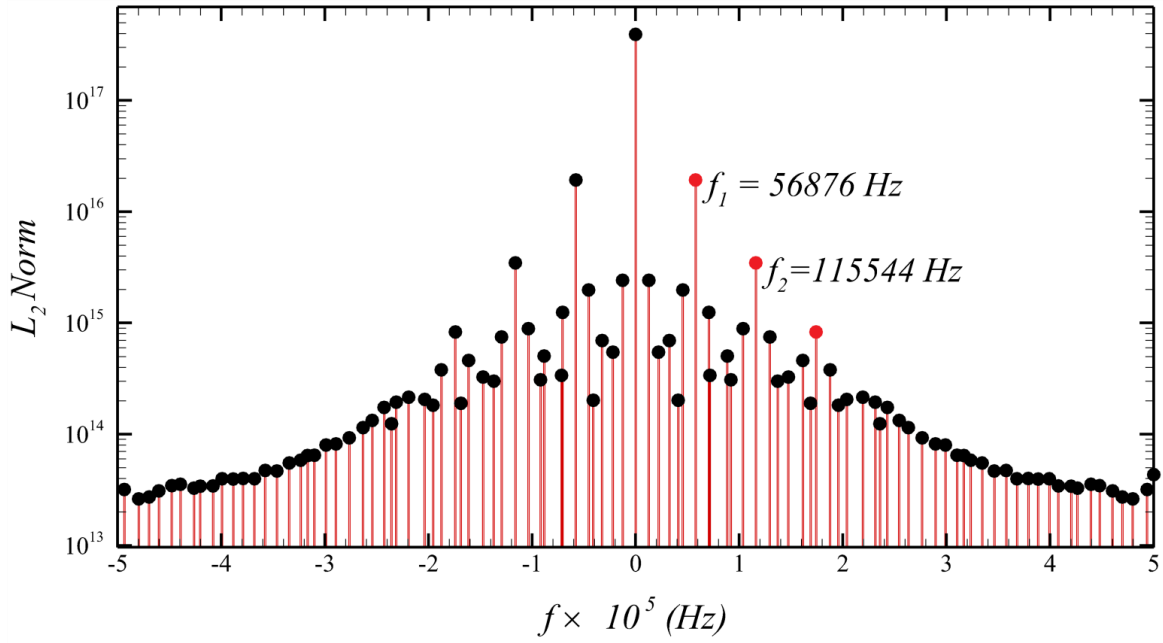

Figure 21: Frequency spectrum of the dynamic modes about the mean mode

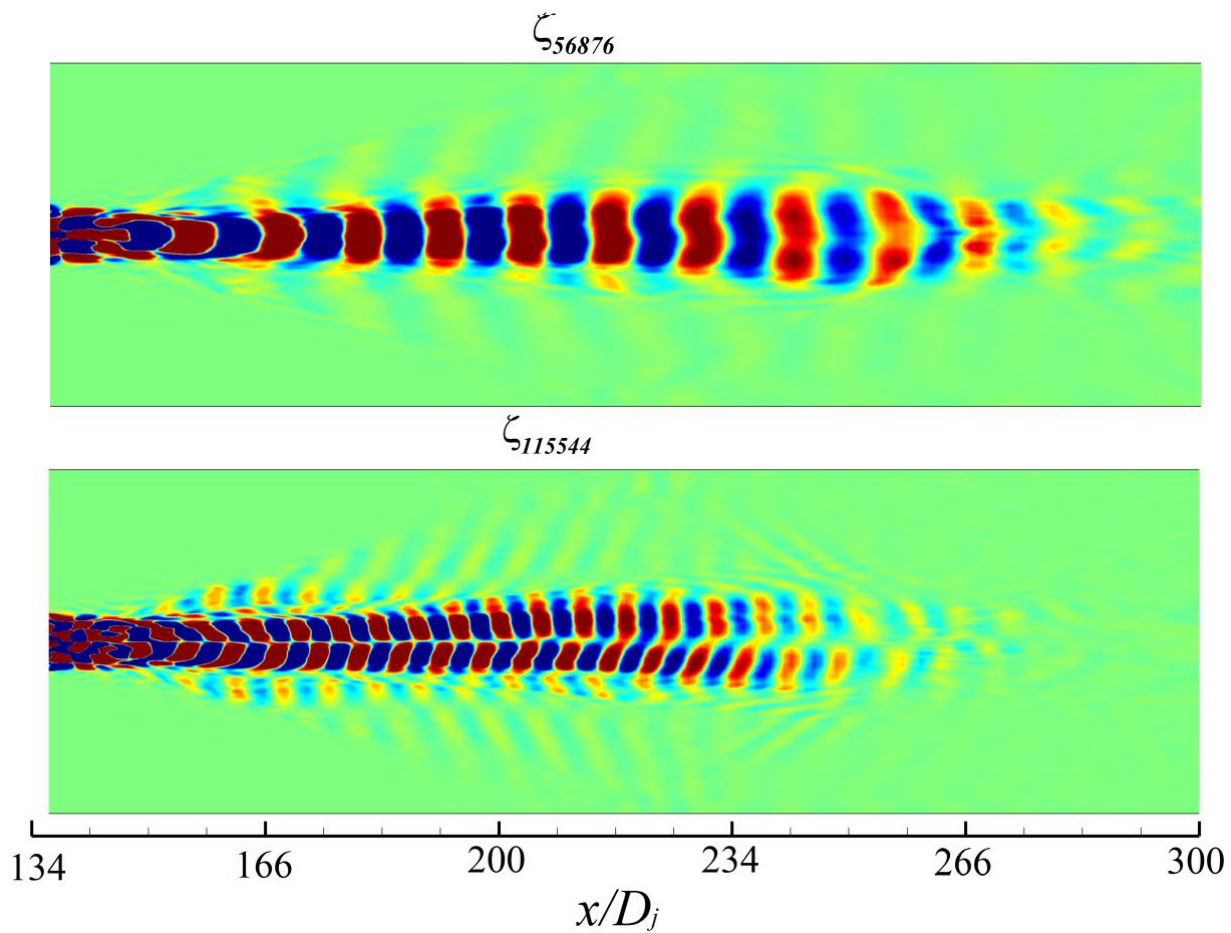

Figure 22: Dynamic modes representing the fundamental and first super harmonic

The real part of the dynamic modes corresponding to $f_1$ and $f_2$, i.e. fundamental and first harmonic computed through the spanwise vorticity is shown in Figure 22. Here $\zeta_{56876}$ and $\zeta_{115544}$ are the dynamic modes where subscript refers the corresponding frequencies. The first dynamic mode related to the fundamental frequency is identical to the $1^{st}$ mode of vorticity based POD (Figure 16). The contour mimics the traveling wave-like motion which physically means downstream advection of the large-scale vortices. The acoustics generated at the shear layer interface is also captured however this acoustics may not play a huge role in modifying turbulence. It is apparent now that the presence of reattachment shock induces instability in the shear layer and alternate shedding of the vortices from the recirculation region induces instability along the shear layer. This instability leads to the breakdown of the jet leading to the formation of the organized structures shedding alternately in a quasi-periodic manner. The initial onset of shear layer rollup is due to the *K-H* instability which undergoes strong gradient upon interacting with the shock induces further instability. This induction of additional instability leads to the interaction of the vortical structures further downstream upon convection and pairing /merging phenomenon leads to the amplification of the scale of the vortices. These vortices upon growing transform into *Von-Karman* street, however, *K-H* type vortices may still be present in the flow. From the contour plot corresponding to the fundamental frequency, one can observe that the vortex grows as it convects downstream which is due to the entrainment of the surrounding flow. At around *x/D$_j$ ≈250* aggressive vortex elongation is revealed, and the quasi two-dimensional vortices seem to exist at this point due to very high strain. As noted earlier it is due to the *shock/vortex interaction* where after passing through the shock wave, the vortex is stretched due to high strain additionally braid region between rollers plays the key role in this mechanism.

The contour corresponding to the first harmonics is identical to the observation of the $3^{rd}$ and $4^{th}$ vorticity POD modes in Figure 16. Since the frequency is approximately double for this mode the wavelength has reduced significantly along with the amplitude of the oscillation. For the dynamic mode related to the fundamental frequency, ten oscillatory periods between *x/D$_j$ ≈140* to *x/D$_j$≈266* are present which convey the advection of the large scale structures, i.e. wake with 2S mode characteristics. However, in case of $\zeta_{115544}$ for the similar interrogation window around nineteen oscillatory periods are present with reduced amplitude and smaller vortices. This is consistent with the observation of the POD decomposition. This points out that the effect of strain in the braid region is manifested into the vortex stretching in the outward region especially in the region between $x/D_j \approx 200$ and $x/D_j \approx 234$. Past this point, the coherence appears to be lost due to the excessive vortex stretching probably leading to the formation of vortex filament elongated in the streamwise direction which is consistent with the results of the instantaneous vorticity contour (Figure 12).

Overall, the combined analysis of instantaneous and decomposed flow field brings out various interesting features and difference of the flow for both the parametric variations, i.e. jet spacing and strut shape (or lip thickness). The strong effect of three-dimensionality is present throughout the flow of the tapered geometry which means the absence of coherent spanwise vortices. Mostly for the $X_2$-$TS(2D_j)$ case (among tapered configurations), few quasi periodic vortical motions are present in the near field which is due to the transverse velocity gradient arising due to the reattachment shock. This high gradient region leads to the breakup of vortices from the existing unstable shear layer; however, for higher spacing (since the gradient is very less), the shear layer manages to recover from the local high gradient due to the distribution of energy in the third dimension. Same is true for the straight strut, though the huge difference in the flow physics is witnessed which suggests that the lip thickness has a significant effect on the mixing characteristics. This difference is mainly due to the high velocity gradient; the presence of stronger reattachment shock close to the jet exit amplifies the vorticity and hence enhances the turbulence which affects the flow favorably. In the light of the current observation, it can be conjectured that local two-dimensionality in the flow can have a humongous effect on the mixing characteristics.

## III. CONCLUSION

In the present study, the effect of jet spacing and strut shape or more precisely the lip thickness is investigated through Large-Eddy Simulation approach. The mixing efficiency and mixing layer thickness exhibit the considerable effect of spanwise spacing between the jets. The density gradient and vorticity contours reveal lack of *K-H* structures for the $X_3(3D_j)$ and $X_5(5D_j)$ cases due to the strong three-dimensionality of the flow field. The presence of the *K-H* vortices locally in $X_2(2D_j)$ case interacts with the streamwise vortices; the ones generated at the strut base due to unsteadiness of recirculation region and improve mixing. Also, it is detected that the wake mode is absent for the higher spacing cases which severely effects the mixing process, whereas for $X_2(2D_j)$ the vortex breakdown near the jet exit leads to the *Von-Karman* street type structures which interact with the transverse vortices and helps to enhance diffusion. The turbulent statistics confirm that the flow in near jet region for the $X_2(2D_j)$ case is mostly two-dimensional due to strong normal velocity gradient, whereas the same is not true for other two cases with higher spacing. Further, the generation of stronger turbulence across the shear layer for lowest spacing due to the interaction of the vortices and strong shock interaction are found to be responsible for the enhanced turbulent mixing. Apart from this observation, it is noticed that irrespective of the similar convective Mach number, $X_2(2D_j)$ case exhibits the presence of two dimensional spanwise rollers which are completely absent for higher spacing. This is in sharp contrast to the popular belief that higher convective Mach number flow field is dominated by the three dimensional structures. However, a most striking difference arises for the straight case with the presence of the large scale, low frequency and more coherent structure in time and space. The presence of *Von-Karman* street vortices in straight configuration has a significant impact on the mixing behaviour. The vorticity generated at the base region for this configuration is higher compared to the tapered case which amplifies the turbulence along the shear layer. There is a significant difference in the strength of the recirculation region which also contributes to the instability of the shear layer due to its unsteady nature. Larger vortices are shed from the recirculation region for straight cases which actually survive many jet diameters while convecting downstream in the straight case. The amplified turbulence and presence of higher velocity gradient play the most critical role in promoting the diffusion across the shear layer. Finally, it can be inferred that with increased lip thickness even for three-dimensional jet leads to the formation of coherent structures locally which have an enormous effect on the flow field.

## ACKNOWLEDGEMENTS

Financial support for this research is provided through IITK-Space Technology Cell (STC). Also, the authors would like to acknowledge the High-Performance Computing (HPC) Facility at IIT Kanpur (www.iitk.ac.in/cc).